\documentclass[a4paper,11pt]{article}
\usepackage{jheppub} 
\pdfoutput=1
\usepackage{graphicx,array}
\usepackage[dvipsnames]{xcolor}
\usepackage{amstext}
\usepackage{amssymb}
\usepackage{slashed}
\usepackage{cancel}
\usepackage{subcaption}
\usepackage{booktabs}
\usepackage{graphicx}
\usepackage{slashed}
\usepackage{subcaption}
\usepackage{physics}
\usepackage{ulem}
\newcommand{\uv}{\text{uv}}
\newcommand{\ir}{\text{ir}}
\newcommand{\horizon}{\text{h}}

\title{ Phase Transition to RS: Cool, not Supercool}

\author{Rashmish K.~Mishra and}
\affiliation{Jefferson Physical Laboratory, Harvard University,\\
17 Oxford Street, Cambridge, MA, 02138, U.S.A.}
\author{Lisa Randall}

\abstract{Motivated by the warped conifold compactification, we model the infrared (IR) dynamics of confining gauge theories in a Randall-Sundrum (RS)-like setup by modifying the stabilizing Goldberger-Wise (GW) potential so that it becomes large (in magnitude) in the IR and back-reacts on the geometry. We study the high-temperature phase by considering a black brane background in which we calculate the entropy and free energy of the strongly back-reacted solution. As with Buchel's result for the conifold~\cite{Buchel:2021yay}, we find a minimum temperature beyond which the black brane phase is thermodynamically unstable. In the context of a phase transition to the confining background, our results suggest that the amount of supercooling that the metastable black brane phase undergoes can be limited. It also suggests the first-order phase transition (and the associated gravitational waves from bubble collision) is not universal. Our results therefore have important phenomenological implications for early universe model building in these scenarios.}

\begin{document} 
\maketitle
\flushbottom

\section{Introduction and Generalities}
It has been argued that the Randall-Sundrum (RS) model undergoes a first-order phase transition from a black brane phase to the RS phase, but only with significant supercooling. This result should be compared to that of Buchel~\cite{Buchel:2021yay}, who studied explicitly the black hole phases in a related warped compactification, where he found a very different result regarding the phase transition from an AdS-Schwarzschild (AdSS)-type unconfined phase to a confined RS-type phase. More explicitly, the focus was on the high-temperature phase of confining gauge theories, modeled in the dual supergravity description, by looking for extended black brane solutions in the warped deformed conifold geometry. His results are that the free energy in the unconfined phase is strongly modified at temperatures below the critical temperature, as compared to the naive expectation from a scaling argument. Eventually, the unconfined phase did not exist beyond a certain temperature, as seen by a divergence in the specific heat and by the speed of sound in the plasma going to zero. This suggests that the warped additional dimensions played an important role, and strongly modify the free energy. Because the warped deformed conifold geometry differs from an RS-like geometry principally in the IR near the would-be horizon, this implies a significant contribution to the free energy in this region. This should be contrasted with the claim of Creminelli et. al.~\cite{Creminelli:2001th} and is presumably associated with the additional light degrees of freedom present in a warped compactification, which are essential in any case to have a full description of the physics contributing to the formation of the brane. In the warped compactification, because of the conifold, a scalar field potential gets large, leading to a different result for the free energy. 

In the scenario considered in~\cite{Buchel:2021yay}, there is an explicit construction of the geometry, including an IR cutoff at the end of the deformed conifold which is in effect the IR brane location at which conformal symmetry is badly broken. The radion determines where the brane forms, but it doesn't in itself provide an explicit construction for how the brane dynamically forms. In general, doing so would require understanding better the dual CFT, and in particular the strong dynamics of the CFT in the IR region.

The calculation in~\cite{Buchel:2021yay} is a 5D calculation, but with seven additional scalars that come from consistent truncation from 10D to 5D (i.e. the 5D equations of motion also solve the 10D equations of motion and the solutions can be uplifted). In our proposed calculation, we aim to understand the IR contribution to quantities such as the free energy and the entropy in a simplified model in which we can better track why the results of the usual RS model and the warped compactification are so different. We do this by including an additional 5D scalar in RS geometry, with a potential that gets large (in magnitude) in the IR, and include its back-reaction on the metric. For the model we choose the large back-reaction survives even though it is happening near a would-be horizon. Furthermore, the normalization of the scalar field will be chosen such that it itself can contribute a classical amount to the free energy (i.e. it is not suppressed by $N^2$) so a full calculation would require computing this contribution as well. 

Let us fix some conventions. We work in mostly positive signature. We set the AdS length scale $\ell_\text{AdS}=1$. Without the scalar sector, the Einstein-Hilbert action is
\begin{align}
    S_\text{GR} = 2M_5^3\int\,\dd^5x\,\sqrt{-g} 
    \left(R-2\Lambda\right)
    + S_\text{bdy}\:,
    \label{eq:GR-action-pureAdS}
\end{align}
where we have included an appropriate boundary term $S_\text{bdy}$ which makes the variation of the bulk action well-defined in the presence of boundaries. The resulting Einstein Equations (EE) are
\begin{align}
    G_{MN} = R_{MN}-\frac12\,R\,g_{MN}=\frac{1}{4M_5^3}T_{MN}=-\Lambda\,g_{MN}\:.
\end{align}
With a negative $\Lambda$, the EE admit a general solution
\begin{align}
    ds^2 = -\rho^2\left(1-\frac{\rho_h^4}{\rho^4}\right)\dd t^2 + \frac{\dd\rho^2}{\rho^2\,\left(1-\frac{\rho_h^4}{\rho^4}\right)} + \rho^2 \dd x^2\:,
\end{align}
where $\rho_h \leq \rho < \infty$, and $\Lambda = -6$ (with $\ell_\text{AdS} = 1$). For applying to the RS setup and getting a normalizable 4D graviton, we need to truncate the $\rho$ coordinate before the asymptotic boundary of AdS at $\rho = \infty$, so the range of coordinates is
\begin{align}
    \rho_h \leq \rho \leq \rho_\text{uv}\:.
\end{align}
There is a UV brane with some tension placed at $\rho=\rho_\text{uv}$. With this range of coordinates, the boundary term in eq.~\eqref{eq:GR-action-pureAdS} becomes relevant---the action has to include an extrinsic curvature term as well as the tension term. These terms cancel on the classical solution, to leading order in $\rho_h^4/\rho_\uv^4$. 

Now we add a 5D scalar $\phi$ to the theory, with some bulk potential $V(\phi)$. The scalar part of the action is
\begin{align}
    S_\phi &= 2 M_5^3\int \dd^5 x\sqrt{-g}\left(-\frac12\,g^{MN}G(\phi)\,\partial_M\phi\,\partial_N\phi - V(\phi)\right)\:,
    \label{eq:scalar_action_after_weyl_rescaling}
\end{align}
where the kinetic term has been kept general. Note that the scalar $\phi$ is dimensionless and is normalized in units of $M_5$. The scalar contributes to the stress-energy tensor as 
\begin{align}
    T_{MN}^\phi = 2M_5^3\,G(\phi)\partial_M\phi\,\partial_N\phi - 2M_5^3\, g_{MN}\left(\frac12G(\phi)(\partial\phi)^2+V(\phi)\right)\:.
\end{align}
For a complete analysis, we need to include the effect of this contribution to the stress-energy tensor on the geometry. 

Before that, we next discuss the scalar potential that can capture the physics in the 5D EFT. For simplicity, in what follows, we will include the cosmological constant term $2\Lambda$ in the scalar potential $V(\phi)$.

\section{Scalar Potential}
The intuition that the result in~\cite{Buchel:2021yay} might be captured by a single 5D scalar $\phi$, with an appropriate potential, comes from the following. In the full geometry, as we go towards the IR, some of the extra dimensions (extra to the fifth) are shrinking. This means that the effective 5D Planck scale decreases, so that the contribution from any potential for $\phi$ should become large w.r.t. the Planck scale in the IR. This means that even as we approach the horizon, the IR contribution cannot necessarily be neglected, as would be true for a constant Planck scale near the horizon.

For a geometry with a black brane, this would imply that as the horizon falls too far beyond the would-be IR brane location, the contribution from the back-reaction of the growing potential could make it thermodynamically unfavorable, or even unstable.

We want a minimal potential to capture this physics. In principle we would have at least two scalars: one to capture the compact directions and one to capture the Goldberger-Wise scalar that is logarithmically running in the noncompact dimension.  Because the radius of the compact dimension is connected to the same flux that determines the Goldberger-Wise scalar, we simplify even further by including a single scalar with the following action in a general Jordan frame.
\begin{align}
    &S = S_\text{GR} + S_\phi + S_\text{bdy}\:,
    \nonumber \\
    &S_\text{GR} = 2M_5^3\int \dd^5x\sqrt{-g}
    \Big(\left(1-\phi/\phi_c\right)^n\,R - 2 \left(1-\phi/\phi_c\right)^m\,\Lambda\Big)\:,
    \nonumber \\
    &S_\phi = 2M_5^3\int \dd^5x\sqrt{-g}
    \Big(-a(\partial\phi)^2 - v(\phi)\Big)\:,
    \label{eq:scalar_action}
\end{align}
where $R$ is the 5D Ricci scalar, $\Lambda$ is the bulk cosmological constant, and we have included a scalar that changes the coefficients of $R$ and $\Lambda$ as it evolves. We assume the scalar has a  bulk potential $v(\phi)$ that gives $\phi$ a non-trivial profile. The scalar $\phi$ is dimensionless and normalized in units of $M_5$. The parameter $a$ is positive, and we have taken a non-canonically normalized kinetic term for $\phi$, for later simplification. Here $m, n, a$ and $\phi_c$ are parameters in the 5D EFT that can be fixed by an explicit 10D example. For $n=m=0,a=1$, we are back to the usual situation that has been studied in the literature before~\cite{Creminelli:2001th}. For simplicity, we take the potential for $\phi$ in the Jordan frame to be just a quadratic
\begin{align}
    v(\phi) = 2\epsilon\phi^2\;,
\end{align}
where we keep in mind that $\epsilon < 0$ for the effect to grow in the IR. As $\phi$ grows in the IR, at some point it approaches $\phi_c$ from below and the effect of $(1-\phi/\phi_c)$ becomes important in the IR causing significant back-reaction to the geometry. This potential is motivated by the shrinking sphere of the conifold in the higher-dimensional theory (e.g. see ref.~\cite{Buchel:2022zxl}), where also the effective five-dimensional Planck scale is reducing which leads to the IR cutoff and before that significant back-reaction in the IR. We will choose parameters in sec.~\ref{sec:results} to focus on the region when the back-reaction is large.

The values of $n,m$ and $a$ are motivated by the warped compactification dynamics we want to model and--when there is a choice--to make the expressions simpler. After a Weyl rescaling of the metric (see app.~\ref{app:pot-in-EF} for details), the scalar part of the action in eq.~\eqref{eq:scalar_action} becomes of the form in eq.~\eqref{eq:scalar_action_after_weyl_rescaling}. For a general $n, m$ in eq.~\eqref{eq:scalar_action}, the functions $G(\phi), V(\phi)$ appearing in eq.~\eqref{eq:scalar_action_after_weyl_rescaling} are given as
\begin{align}
    G(\phi) &= 2a\left(1-\frac{\phi}{\phi_c}\right)^{-n} + \frac{8n^2}{3\phi_c^2}\left(1-\frac{\phi}{\phi_c}\right)^{-2}\:,
    \nonumber \\
    V(\phi) &= 2\epsilon \phi^2 \left(1-\frac{\phi}{\phi_c}\right)^{-\frac{5n}{3}} 
    + 2\Lambda\left(1-\frac{\phi}{\phi_c}\right)^{-\frac{5n}{3}+m}\:.
    \label{eq:G-and-V-generic}
\end{align}
We choose $m=5n/3$ so that setting $\epsilon = 0$ decouples the effect of $\phi_c$ in the potential (in the Einstein frame), and we recover the potential of pure RS, with only a bulk cosmological constant. We also choose $n=2$ and $a=1$ to make $G(\phi)$ simpler. We note however that the methods in our work can be applied in a straightforward manner to general values of $m, n$. With this, the $G(\phi)$ and $V(\phi)$ that we work with in the rest of the paper are
\begin{align}
    G(\phi) &= \left(2+\frac{32}{3\phi_c^2}\right)\left(1-\frac{\phi}{\phi_c}\right)^{-2}\:,
    \nonumber \\
    V(\phi) &= 2\epsilon \phi^2 \left(1-\frac{\phi}{\phi_c}\right)^{-10/3} 
    + 2\Lambda\:.
    \label{eq:G-and-V-specific}
\end{align}
In the limit of $\phi \ll \phi_c$, both $G(\phi)$ and $V(\phi)$ simplify as expected and this limit corresponds to a scalar with just a bulk mass term. Note that for any general polynomial potential $v(\phi)$ in the Jordan frame (in eq.~\eqref{eq:scalar_action}), the potential $V(\phi)$ in the Einstein frame (in eq.~\eqref{eq:scalar_action_after_weyl_rescaling}) gets a factor of $(1-\phi/\phi_c)^{-10/3}$ from the Weyl rescaling, and this gives the large growth behavior as $\phi$ approaches $\phi_c$ in the IR. The kinetic term also gets large as $\phi$ approaches $\phi_c$. While we will work with this non-canonical basis in the numerical work presented later, it is also useful to look at the potential in the canonical basis for the kinetic term. Defining $\sigma$ related to $\phi$ as
\begin{align}
    \frac{\sigma}{\sigma_c} &= -\log \left(1-\frac{\phi}{\phi_c}\right)\:,\:\: \sigma_c = \left(\frac{32}{3\phi_c^2}+2\right)^{1/2}\,\phi_c\:,
    \label{eq:phi-vs-sigma}
\end{align}
the kinetic term for $\sigma$ is canonical. Since $1-\phi/\phi_c$ is always smaller than 1, $\sigma >0$, and as $\phi\to\phi_c$, $\sigma\to\infty$. The potential as a function of $\sigma$ is given as
\begin{align}
    V(\sigma) &= 2\Lambda + 2\epsilon\phi_c^2\:e^{\frac{10}{3}\frac{\sigma}{\sigma_c}}\left(1-e^{-\frac{\sigma}{\sigma_c}}\right)^2\:
    \nonumber \\
    &= 2\Lambda + 2\tilde{\epsilon}\sigma_c^2\:e^{\frac{10}{3}\frac{\sigma}{\sigma_c}}\left(1-e^{-\frac{\sigma}{\sigma_c}}\right)^2\:,
    \label{eq:potential_sigma}
\end{align}
where we have defined $\tilde{\epsilon} = \epsilon (\phi_c/\sigma_c)^2$ in the second line, for notational simplicity.
This potential depends exponentially on $\sigma$, and for $\sigma/\sigma_c \gg 1$, it becomes purely exponential: 
\begin{align}
    V(\sigma) \:\:\underset{\sigma \gg \sigma_c}{ = }\:\: 2\tilde{\epsilon}\sigma_c^2 e^{\frac{10}{3}\frac{\sigma}{\sigma_c}}\:.
    \label{eq:potential_sigma_large}
\end{align}
The exponential dependence of the potential on the canonically normalized field is quite generic in dimensional reductions, and should not be a surprise. Even though the potential we work with is a pure exponential only for large field values, it is useful to keep this limit in mind because analytical solutions are known for a purely exponential scalar potential and their adiabatic generalizations~\cite{Chamblin:1999ya, Gubser:2008ny}. We will use them to understand some of the numerical results we obtain, and we will have more to say about them in later sections.

It is also useful to look at the potential in the other limit of $\sigma \ll \sigma_c$, where it looks like
\begin{align}
    V(\sigma) \:\:\underset{\sigma \ll \sigma_c}{ = }\:\: 2\Lambda + 2\tilde{\epsilon}\sigma^2 
    + \frac{14}{3}\frac{\tilde{\epsilon}}{\sigma_c}\,\sigma^3
    +\frac{101}{18}\frac{\tilde{\epsilon}}{\sigma_c^2}\,\sigma^4+\cdots \qquad  ,
    \label{eq:potential_sigma_small}
\end{align}
and one generically gets cubic and higher-order self-interaction terms. If the scalar $\sigma$ was interpreted as a Goldberger-Wise (GW) scalar, we see that self-interactions in the potential are necessary to capture the IR physics of this model correctly~\cite{Chacko:2013dra, Mishra:2023kiu}, and their exact form is fixed by a specific UV completion.
\begin{figure}
    \centering
    \includegraphics[width=\textwidth]{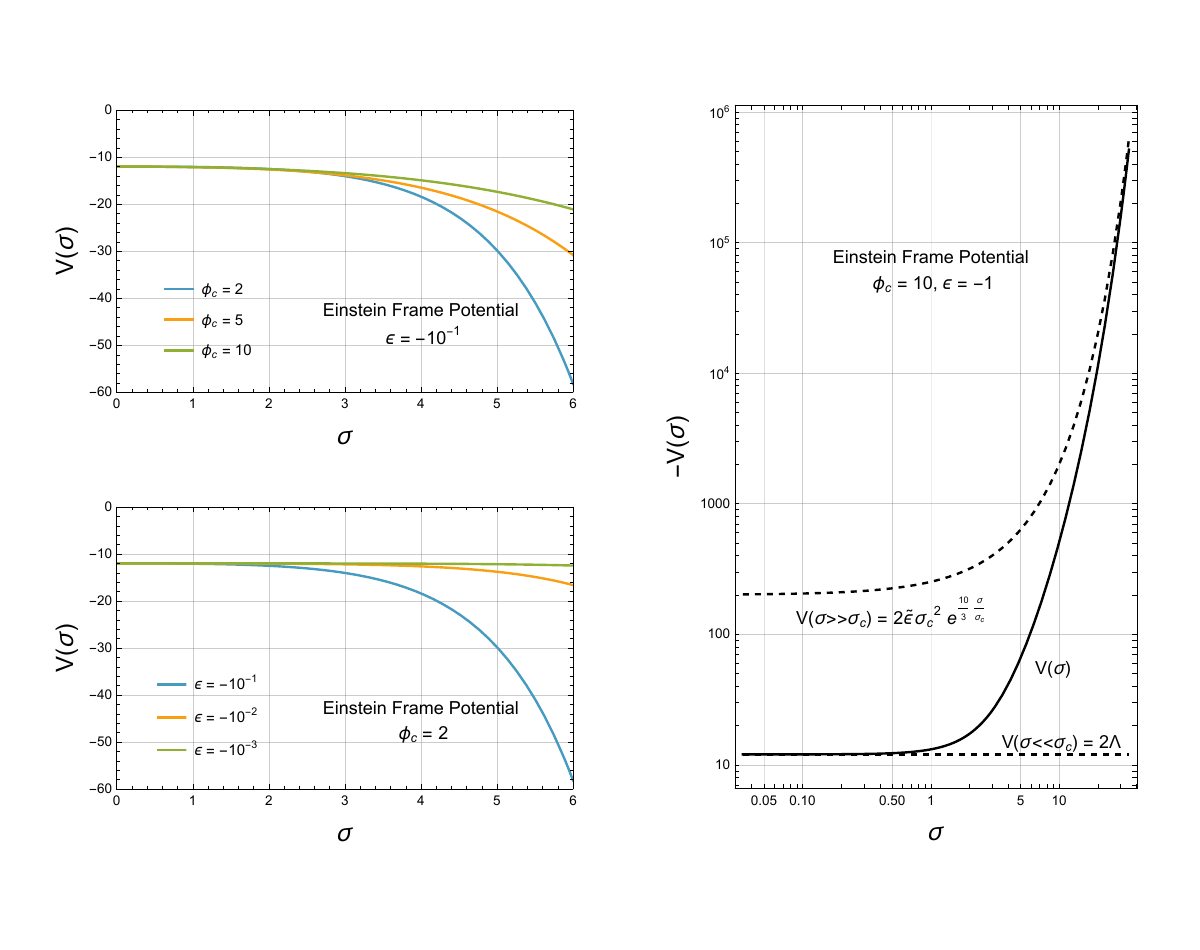}
    \caption{\small{The Einstein frame potential $V(\sigma)$ for the canonically normalized scalar $\sigma$: as $\phi_c$ (left, top) and $\epsilon$ (left, bottom) is varied. Right plot shows the negative of the potential (in solid, for $\phi_c=10, \epsilon=-1$) sandwiched between the small $\sigma$ asymptote of $2\Lambda$ and the large $\sigma$ asymptote of $2\widetilde{\epsilon}\sigma_c^2\,e^{10\sigma/(3\sigma_c)}$. }}
    \label{fig:potential_caseAandB}
\end{figure}

Fig.~\ref{fig:potential_caseAandB} shows the potential $V(\sigma)$ for some choice of parameters and compares it to the asymptotic forms. The common feature is that the potential starts from the negative value of $2\Lambda$ (the cosmological constant in AdS), grows more negative as the field value increases, and eventually asymptotes to an exponential. The changes in the behavior are controlled by $\epsilon$ and $\phi_c$. Note that an exponentially large and negative potential is bounded from above and does not lead to any pathologies, as argued in~\cite{Gubser:2000nd}, based on allowing ``good'' singularities that have a field theory interpretation.

\section{Coordinates and Coupled Dynamics}
Given the potential $V(\phi)$, we can solve for the background $\phi$ profile and the background metric containing a black brane taking back-reaction into account. This geometry is dual to the deconfined phase. We choose coordinates and the relevant boundary conditions to fully specify the system of equations and obtain the solution. Notice that most work has focused on modifications to the free energy of the confined phase. Here we show that generically we can expect a big back-reaction on the deconfined phase as well. The strong modification in the free energy will change the critical temperature $T_c$ and we expect it to be of the same order as the confinement scale (unlike the small back-reaction case in~\cite{Creminelli:2001th}).

Our goal is to calculate the free energy of the deconfined phase. For this purpose, it is useful to work in the Euclidean version of the theory, with the Euclidean time compactified on a circle of circumference $\beta = 1/T$ where $T$ is the temperature of the deconfined phase. Recall that the Euclidean metric with a black brane, without any scalars in the theory, is $(\ell_\text{AdS} = 1$)
\begin{align}
    ds^2 = \rho^2\left(1-\frac{\rho_h^4}{\rho^4}\right)\dd t_E^2 + \frac{\dd\rho^2}{\rho^2\,\left(1-\frac{\rho_h^4}{\rho^4}\right)} + \rho^2 \,\dd x^2\:,\:\: \rho_h \leq \rho \leq \rho_\uv\:,
    \label{eq:AdSS_metric}
\end{align}
where the horizon (extended in the transverse directions) is at $\rho=\rho_h$ and the space is truncated at $\rho = \rho_\uv$, before the asymptotic AdS boundary at $\rho\to\infty$. Given the location of the horizon $\rho_h$, the periodicity of the time circle is fixed for the Euclidean geometry to be smooth at the horizon where the time circle shrinks to zero size. We will have more to say about this in the next section.

A general parameterization of the black brane metric allowing for  back-reactions is
\begin{align}
    ds^2 = a^2(\rho)\rho^2\left(1-\frac{\rho_h^4}{\rho^4}\right)\dd t_E^2 + \frac{b^2(\rho)\dd\rho^2}{\rho^2\,\left(1-\frac{\rho_h^4}{\rho^4}\right)} + c^2(\rho)\rho^2\, \dd x^2\:.
\end{align}
We can define $\rho$ such that $c(\rho)=1$. For numerical convenience, we define a new coordinate
\begin{align}
    \xi = \frac{\rho^4 -\rho_h^4}{\rho_\text{UV}^4-\rho_h^4}\:,\qquad 0\leq\xi\leq1\:.
    \label{eq:xi-coordinate}
\end{align}
In terms of $\xi$, the horizon is located at $\xi=0$ and the UV boundary is located at $\xi=1$. In $\xi$ coordinates the pure black brane metric (without  scalars--eq.~\eqref{eq:AdSS_metric}) looks like
\begin{align}
    \dd s^2 = \frac{\xi\,\rho_\horizon^2\alpha^{-1/2}}{\sqrt{\xi+\alpha}}\,\dd t_E^2 + \frac{1}{16\xi(\xi+\alpha)}\,\dd\xi^2+\sqrt{\xi+\alpha}\,\rho_\horizon^2\alpha^{-1/2}\,\dd x^2\:,
    \label{eq:unrescaled-T-Vol3}
\end{align}
where $\alpha = \rho_\horizon^4/(\rho_\uv^4-\rho_\horizon^4) \approx (\rho_h/\rho_\uv)^4$ for $\rho_h \ll \rho_\uv$. Rescaling $t_E,x$ as 
\begin{align}
    t_E \to \rho_h \alpha^{-1/4}\,  t_E\, ,  \qquad x \to \rho_h \alpha^{-1/4} \, x \:,
\end{align}
the metric becomes
\begin{align}
    \dd s^2 = \frac{\xi}{\sqrt{\xi+\alpha}}\,\dd t_E^2 + \frac{1}{16\xi(\xi+\alpha)}\,\dd\xi^2+\sqrt{\xi+\alpha}\,\dd x^2\:.
    \label{eq:rescaled-T-Vol3}
\end{align}
This rescaling, which is a large diffeomorphism, is important when calculating quantities per unit volume, and when the periodicity of the Euclidean time circle is identified with the temperature. To be more explicit, the 3-volume and the temperature in the rescaled coordinates (labeled by a superscript $\xi$) are related to the non-rescaled ones as
\begin{align}
    \text{Vol}_3^\xi &= \frac{\rho_h^3}{\alpha^{3/4}}\,\text{Vol}_3 = \frac{\rho_\uv^3}{\left(1+\alpha\right)^{3/4}}\,\text{Vol}_3 \underset{\alpha \ll 1}{\approx} \rho_\uv^3 \text{Vol}_3\:,
    \nonumber \\
    T^\xi &=\frac{\alpha^{1/4}}{\rho_h}\,T = \frac{\left(1+\alpha\right)^{1/4}}{\rho_\uv}T
    \underset{\alpha \ll 1}{\approx} \frac{T}{\rho_\uv}\:.
    \label{eq:T-vs-Txi-Vol-vs-Volxi}
\end{align}
We can see that in the near horizon limit, $\xi\to 0$, the $(tt)$ component of the metric vanishes, and the $(\xi\xi)$ component blows up, as expected in a background with a horizon. We can now parameterize the back-reacted black brane background metric to be
\begin{align}
    \dd s^2 = a^2(\xi)\,\xi\,\dd t_E^2 + \frac{b^2(\xi)}{\xi}\,\dd\xi^2+\sqrt{\xi+\alpha}\,\dd x^2\:,
    \label{eq:AdSS-metric-ansatz}
\end{align}
where we have kept the coefficient of $\dd x^2$ unchanged, since that can be absorbed in the definition of $x$. We can read off the functional form of $a(\xi), b(\xi)$ in the limit of no back-reaction:
\begin{align}
    a(\xi) \:\:&\underset{\text{no b.r.}}{=}\:\: \frac{1}{\left(\xi+\alpha\right)^{1/4}}\:,
    \nonumber \\
    b(\xi) \:\:&\underset{\text{no b.r.}}{=}\:\: \frac{1}{4\left(\xi+\alpha\right)^{1/2}}\:.
    \label{eq:a_b_no_br}
\end{align}
For completeness, we also note the RS metric (dual to the confined phase) in the $\xi$ coordinates. In $\rho$ coordinates, the RS metric is obtained from the black brane metric by setting $\rho_h = 0$ and truncating the spacetime at $\rho = \rho_\ir$ instead. In $\xi$ coordinate, one can set $\rho_h = 0$ after the $\rho_h$ dependent rescaling of $t, x$ has been performed (same as going from eq.~\eqref{eq:unrescaled-T-Vol3} to eq.~\eqref{eq:rescaled-T-Vol3}). Starting with the RS metric in $\rho$ coordinates
\begin{align}
    \dd s^2 = \rho^2\,\dd t_E^2 + \frac{\dd\rho^2}{\rho^2}+ \rho^2\,\dd x^2\:,
    \qquad \rho_\ir \leq \rho \leq \rho_\uv\:,
\end{align}
The metric in $\xi$ coordinates, after $t, x$ rescaling, looks like
\begin{align}
    \dd s^2 = \sqrt{\xi+\alpha}\,\dd t_E^2 + \frac{\dd\xi^2}{16\,(\xi+\alpha)^2} + \sqrt{\xi+\alpha}\,\dd x^2\:,
    \qquad \xi\ir \leq \xi \leq 1\:,
\end{align}
where $\xi_\ir = (\rho_\ir - \rho_h)/(\rho_\uv - \rho_h)$. Note that unlike the black-brane case, the lower limit of $\xi$ is not zero anymore, and depends on the relation between $\rho_\ir$ and $\rho_h$.  Also, $\alpha = \rho_h^4/(\rho_\uv^4 - \rho_h^4)$ appearing in the metric is simply to be taken as a parameter, since there is no horizon in the RS metric at $\rho = \rho_h$. This choice allows using the $\xi$ coordinate (defined in eq.~\eqref{eq:xi-coordinate}) for both the RS and the black-brane spacetime. The functional form of $a(\xi), b(\xi)$ for the RS metric, in the limit of no back-reaction, can be read off easily.

\subsection{Coupled Equations}
At this point, we have three scalar functions to solve for: $a(\xi), b(\xi)$ and $\phi(\xi)$. Using the ansatz for the black brane metric, eq.~\eqref{eq:AdSS-metric-ansatz}, the $\xi\xi$ component of the EE turns out to be an algebraic equation in $b(\xi)$, and it can be solved in terms of $a(\xi), \phi(\xi)$ and its derivatives.\footnote{The $tt$ component of the Einstein equations is linearly dependent on the other equations, due to the Bianchi identity.} This constraint is given as (note that the potential $V(\phi)$ includes the cosmological constant)
\begin{align}
    \frac{1}{\xi}\,V(\phi)\,b^2(\xi) &= \frac12G(\phi)\phi'^2 
    -\frac{3}{2(\xi+\alpha)}\frac{a'}{a}
    -\frac{1}{(\xi+\alpha)^2}
    \left(
    \frac{9}{8}+\frac{3\alpha}{4\xi}
    \right)\:.
    \label{eq:b-eqn-constraint}
\end{align}
In the limit of no back-reaction, setting $\phi=0, \phi'=0, V=2\Lambda, \Lambda = -6$ and $a(\xi) = (\xi+\alpha)^{-1/4}$, the above reduces to $1/b(\xi) = 4(\xi+\alpha)^{1/2}$, in agreement with eq.~\eqref{eq:a_b_no_br}.

The $ii$ components of the EE are non-algebraic and yield the same equation due to the rotational symmetry of the spacetime:
\begin{align}
    \left(2\alpha +\frac34\xi
    \right)
    +
    \frac{a'}{a}
    \left(6\alpha^2 +14\alpha\xi +8\xi^2 
    \right)
    -
    \frac{b'}{b}
    \left(2\alpha^2 +6\alpha\xi +4\xi^2 
    \right)
    \nonumber \\
    +
    \left(\frac{a''}{a}-\frac{a'}{a}\frac{b'}{b}\right)
    \left(4\alpha^2\xi +8\alpha\xi^2 +4\xi^3 
    \right)
   +(\alpha+\xi)^2\Big(2b^2V+\xi G \phi'^2\Big) = 0\:.
   \label{eq:EOM-GR}
\end{align}
Putting the $b(\xi)$ constraint in the above gives an equation in terms of $a,\phi$ and their derivatives. The explicit equation is long and does not give any insights. 

The scalar equation is
\begin{align}
    & \phi '' + \phi '\left(\frac{a'}{a} - \frac{b'}{b}+\frac{4 \alpha +7 \xi}{4\xi(\alpha +\xi)} \right)
    + \frac12 \phi '^2 \frac{G'}{G}
    -\frac{b^2}{\xi} \frac{V'}{G} =0\:.
    \label{eq:EOM-scalar}
\end{align}
We have again left it in terms of $b,b'$. In the limit of no back-reaction, using $V(\phi) = 2\epsilon\phi^2 + 2\Lambda, \Lambda = -6$, $G(\phi) = 1$, and setting $4b(\xi) = (\xi+\alpha)^{-1/2}$, $a(\xi) = (\xi+\alpha)^{-1/4}$, the scalar equation reduces to
\begin{align}
    \xi(\alpha+\xi)\phi''+(\alpha+2\xi)\phi'-\frac{\epsilon}{4}\phi =0\:,
\end{align}
which matches the equation of motion for a scalar with a quadratic bulk potential~\cite{Creminelli:2001th} in the limit of no back-reaction.\footnote{There is a subtlety when taking the limit of no back-reaction for the $\phi$ equations of motion---the constraint equation for $b(\xi)$ depends on $\phi(\xi)$, so that one must set $\phi = 0$ in the constraint equation to recover $4b(\xi) = (\xi+\alpha)^{-1/2}$. If the constraint equation is already used in the $\phi$ equation, taking the limit of small back-reaction is more involved.} There are two solutions to the second-order differential equation that are known analytically. Requiring regularity at the horizon ($\xi=0$), the regular solution is
\begin{align}
    \frac{\phi(\xi)}{\mathcal{N}} = \:  _2F_1\left(\frac{1}{2}-\frac{\sqrt{\epsilon +1}}{2},\frac{\sqrt{\epsilon +1}}{2}+\frac{1}{2};1;-\frac{\xi }{\alpha }\right)\:,
    \label{eq:phi-sol-quadratic-potential}
\end{align}
where $\mathcal{N}$ is chosen such that $\phi(\xi=1) = \phi_\uv$. We will use this solution as a seed for solving the coupled differential equations numerically. 

\subsection{Boundary Conditions}
\label{subsec:boundary-conditions}
The equations of motion reduce to two second-order differential equations, eq.~\eqref{eq:EOM-GR} and eq.~\eqref{eq:EOM-scalar}. We need four boundary conditions in order to obtain the solutions. The two UV boundary conditions are straightforward, whereas the two IR boundary conditions are ``regularity'' at the horizon $\xi=0$, i.e. the functions $\phi, a$ and its derivatives should be bounded in magnitude. In the absence of an analytical solution, it is not straightforward to impose this. We now discuss the strategy to impose this condition numerically. Ultimately we will choose UV boundary conditions that guarantee regularity at the horizon. 

A function that is regular at the horizon can be expanded in a power series around the location of the horizon, with no singular terms. This implies
\begin{align}
    a(\xi) &\underset{\xi\to 0}{=} a_0 + a_1 \xi +  \cdots\:,
    \nonumber \\
    \phi(\xi) &\underset{\xi\to 0}{=} \phi_0 + \phi_1 \xi + \cdots\:.
\end{align}
Further, since these must be solutions to the equations of motion, we can plug these expansions into the equations of motion to solve for the coefficients recursively. Given $(a_0, \phi_0)$, this fixes $a_1, \phi_1$ uniquely. Generalizing this procedure, given the coefficients to a given order, the coefficients in the next order are fixed. We will work here with only the first order, for simplicity. Our results can be improved by a higher-order analysis (in app.~\ref{app:numericalMethod-Alternate} we provide an alternate numerical method to generate these solutions). We therefore have 
\begin{align}
    a_1 = a_1(a_0, \phi_0)\:,\:\: \phi_1 = \phi_1(a_0, \phi_0)\:\:, 
\end{align}
in which the exact functional form depends on $G(\phi)$ and $V(\phi)$. An explicit expression for them is given in app.~\ref{app:IR-derivatives}.

We now have a numerical strategy for getting a regular solution. We can solve the second-order coupled differential equations for a given $a(\xi_\uv), \phi(\xi_\uv), a'(\xi_\uv), \phi'(\xi_\uv)$. From this we can read the values $a(\xi_\ir), \phi(\xi_\ir), a'(\xi_\ir), \phi'(\xi_\ir)$ for some $\xi=\xi_\ir < \xi_\uv$. We choose $\xi_\ir \ll 1$ to be very close to the horizon $\xi=0$ (in a way that 
will be made more precise later). We then have two different expressions for $\phi(\xi)$ in the vicinity of $\xi=\xi_\ir$, depending on whether $\xi > \xi_\ir$ or $\xi < \xi_\ir$:
\begin{align}
    \xi > \xi_\ir\::&\:\: \phi(\xi) = \phi(\xi_\ir) + \phi'(\xi_\ir)(\xi-\xi_\ir) + \cdots\:,
    \nonumber \\
    \xi < \xi_\ir\::&\:\: \phi(\xi) = \phi_0 + \phi_1(a_0, \phi_0) \xi\: + \cdots.
\end{align}
Matching the values and the first derivative at $\xi = \xi_\ir$ gives
\begin{align}
    \phi'(\xi_\ir) &= \phi_1(a_0, \phi_0)\:,
    \nonumber \\
    \phi(\xi_\ir) &= \phi_0 + \phi_1(a_0, \phi_0)\xi_\ir \approx \phi_0\:.
\end{align}
We will try to find a regime when the approximation in the last line is justified. This requires choosing a very small $\xi_\ir$, for finite $\phi_1$ (which must hold for a regular solution). The same steps can be argued for $a(\xi)$ as well. Hence for a regular solution, we require
\begin{align}
    \phi'(\xi_\ir) &= \phi_1 (a(\xi_\ir), \phi(\xi_\ir))\:,
    \nonumber \\
    a'(\xi_\ir) &= a_1 (a(\xi_\ir), \phi(\xi_\ir))\:.
    \label{eq:phi-prime-vs-phi1}
\end{align}
Since $\phi_1$ and $a_1$ are known functions of their arguments given the choice of potentials, eq.~\eqref{eq:phi-prime-vs-phi1} above gives two non-trivial conditions. We then vary two of the four UV boundary conditions until these conditions are satisfied. We choose to vary $a'(\xi_\uv)$ and $\phi'(\xi_\uv)$. The value $a(\xi_\uv)$ is chosen to be the same as the no back-reaction case: $a(\xi_\uv) = \left(\xi_\uv+\alpha\right)^{-1/4}$ (see eq.~\eqref{eq:a_b_no_br}), while $\phi(\xi_\uv) = \phi_\uv$ is a free parameter, which we choose so that it can yield a non-trivial back-reaction.  Further, for simplicity we choose $\xi_\uv = 1$.

We next need to understand what value of $\xi_\ir$, which should be ``close'' to the horizon $\xi=0$, will be ``small enough'' to justify our procedure. For this, it is instructive to look at the functional form of $a(\xi)$ when there is no back-reaction: $a(\xi) = \left(\xi+\alpha\right)^{-1/4}$. When $\xi \gg \alpha$, there is no effect of  $\alpha$, i.e. the effect of the horizon is not seen, while once $\xi \lesssim \alpha$, the function is dominated by the horizon contribution. This suggests that when $\xi\ll\alpha$, we are close to the horizon. This is more clearly seen in the proper distance from the point $\xi=\xi_\ir$ to the horizon $\xi=0$, which is $(1/2)\sinh^{-1}({\sqrt{\xi_\ir/\alpha}})$. For a fixed and small proper distance, we need $\xi_\ir/\alpha$ to be fixed and small. Note that $\alpha \approx (\rho_h/\rho_\uv)^4 \ll 1$ already, so this requires $\xi_\ir$ to be even smaller. We will choose to work with $\xi_\ir/\alpha = 10^{-2}$  in our numerical results.

\section{Thermodynamic Quantities in the Deconfined Phase}
\label{sec:FE-and-Temperature}
We would like to calculate the free energy and entropy of the deconfined phase as a function of temperature, including the effect of the scalar. In our model, the scalar should mimic the effect of the shrinking conifold and therefore has a potential that gets large in the IR. In the dual 5D picture in the semi-classical limit, we need to solve for the metric and the scalar profile. We vary $\alpha \approx (\rho_h/\rho_\uv)^4$, and solve for the metric and the scalar profile as a function of $\alpha$. From these profiles, we calculate the temperature, the entropy and the free energy as a function of $\alpha$. When the back-reaction is small, smaller $\alpha$ corresponds to a larger proper distance between the horizon and the UV brane, and a smaller temperature. This relation however does not hold when the back-reaction is large. In the subsections below, we clarify the meaning of changing $\alpha$. Given the temperature, the entropy and the free energy as a function of $\alpha$ we can obtain the entropy and free energy implicitly as a function of temperature. There are multiple ways to calculate the temperature, the entropy, and the free energy. We use the most convenient one, which we explain in the subsections below. All of these quantities will change significantly from their values in a pure gravitational solution, due to the back-reaction. 

\subsection{Entropy and the Meaning of Changing $\alpha$}
\label{subsec:Entropy}
The entropy of the deconfined phase can be calculated by the Bekenstein-Hawking formula for the black brane geometry. Since the horizon is extended in the $\vec{x}$ directions, we expect the entropy to scale with the three-volume. We denote the extensive and intensive quantities by upper and lower case letters, in what follows. We therefore have 
\begin{align}
    s &= \frac{S}{\text{Vol}_3} = \frac{\mathcal{A}_\text{horizon}}{4 G_N} \frac{1}{\text{Vol}_3}\:.
\end{align}
Using the metric for the black brane geometry (see eq.~\eqref{eq:AdSS-metric-ansatz}), we can calculate the ``area'' of the codimension-2 horizon located at $\xi=0$,
\begin{align}
    \mathcal{A}_\text{horizon} & = \alpha^{3/4}\,\text{Vol}_3^\xi = \frac{\alpha^{3/4}}{\left(1+\alpha\right)^{3/4}}\,\rho_\uv^3\,\text{Vol}_3\:,
\end{align}
while in the normalization of the action we are using (see eq.~\eqref{eq:GR-action-pureAdS}), $1/G_N = 32\pi M_5^3 $, using which we get
\begin{align}
    s &= 8\pi M_5^3 \frac{\alpha^{3/4}}{\left(1+\alpha\right)^{3/4}}\rho_\uv^3\:.
    \label{eq:entropy-BH}
\end{align}
Due to the parameterization chosen for the metric, the formula for the entropy is very simple and does not require any further computation. The above equation holds both at small and large back-reaction, because the area of the horizon does not depend on the scalar functions $a(\xi), b(\xi)$, in the coordinate system we have used (see eq.~\eqref{eq:AdSS-metric-ansatz}). This relation between the entropy and $\alpha$ clarifies the meaning of changing $\alpha$: a smaller $\alpha$ corresponds to a smaller entropy density. When the back-reaction is small, smaller $\alpha$ means smaller temperature. As we will see later in the section, the relation between temperature and $\alpha$ is very different when there is strong back-reaction.\footnote{The relation between $\alpha$ and the proper distance of the horizon from the UV also gets modified---smaller $\alpha$ does not correspond to a larger proper distance of the horizon from UV, unlike the no back-reaction case. As discussed in sec.~\ref{subsec:proper-distance-of-horizon}, the proper distance saturates for smaller $\alpha$.} Changing $\alpha$ is therefore the natural continuation of changing temperature, when the back-reaction is strong. This is the approach we take here to generate the phase curves. In the limit of no back-reaction, using the expressions for $a(\xi), b(\xi)$ from eq.~\eqref{eq:a_b_no_br} in eq.~\eqref{eq:Temperature-From-a-b-functions} we get $\pi T/\rho_\uv = \alpha^{1/4}/(1+\alpha)^{1/4}$, using which, the entropy in eq.~\eqref{eq:entropy-BH} is $s = 8\pi^4 M_5^3 T^3$. This is the expected result in the limit of no back-reaction.

\subsection{Temperature}
\label{subsec:Temperature}
Correlators of the field theory at finite temperature can be calculated from the thermal partition function, which is obtained from the path integral by making the time direction Euclidean and compactifying on a circle of radius $\beta = 1/T$. In the gravitational description, the partition function is calculated in the semi-classical limit by considering solutions to the equations of motion that have the correct asymptotic behavior. In the absence of any localized source for stress-energy tensor, the solution must be smooth everywhere as well. This latter requirement fixes the temperature $T$ in terms of the geometrical parameters. To ensure smoothness for the black brane (eq.~\eqref{eq:AdSS-metric-ansatz}) in the Euclidean signature and close to the horizon location, we restrict to the $(t_E,\xi)$ plane in the vicinity of the horizon $\xi\to0$. The metric in this vicinity looks locally as
\begin{align}
    \dd s^2\underset{\xi\to0}{=} a^2(0)\,\xi\, \dd t_E^2 + b^2(0)\,\xi^{-1}\,\dd\xi^2\:,
\end{align}
where we have assumed that $a, b$ have a constant value as $\xi\to0$. This is certainly true when there is no back-reaction, and is expected in general for a regular solution.

By a coordinate change from $(\xi,t_E)\to(r,\theta)$, the metric in the vicinity of the horizon can be put in a more familiar form. Defining the coordinate $r$ by the equation $b(0) d\xi/\sqrt{\xi} = dr$, we get $\xi = (r/2b(0))^2$, where the integration constant is chosen so that $r = 0$ at $\xi = 0$. Recall that the coordinate $t_E$ is periodic with fundamental period $(0, 1/T^\xi)$, since $t_E$ was rescaled in going from eq.~\eqref{eq:unrescaled-T-Vol3} to eq.~\eqref{eq:rescaled-T-Vol3} and had a fundamental period $(0, 1/T)$ before the rescaling (see eq.~\eqref{eq:T-vs-Txi-Vol-vs-Volxi} for the relation between $T$ and $T^\xi$). We define $\theta = 2\pi T^\xi t_E$ so that $\theta$ is periodic, with fundamental period $(0,2\pi)$, given that $t_E$ is periodic with fundamental period $(0,1/T^\xi)$. In terms of $(r,\theta)$ the metric near the horizon becomes
\begin{align}
    \dd s^2 \underset{\xi\to0}{=} \:\:\frac{a^2(0)}{4b^2(0)}\frac{1}{(2\pi T^\xi)^2}\,r^2\,\dd\theta^2 + \dd r^2\:.
\end{align}
Requiring the region $\xi\to0$ to be without any conical singularities, the metric must reduce to the metric of flat space in polar coordinates. This is ensured if
\begin{align}
    \frac{a^2(0)}{4b^2(0)}\frac{1}{(2\pi T^\xi)^2}  = 1\:.
\end{align}
Using $T^\xi = (\alpha^{1/4}/\rho_h) T$ from eq.~\eqref{eq:T-vs-Txi-Vol-vs-Volxi}, the temperature is given in terms of $a, b$ functions as
\begin{align}
    T &= \underset{\xi\to0}{\text{lim}}\:\:\frac{\rho_h}{\alpha^{1/4}} \frac{a(\xi)}{4\pi b(\xi)}
     = \underset{\xi\to0}{\text{lim}}\:\:\frac{\rho_\uv}{\left(1+\alpha\right)^{1/4}} \,\frac{a(\xi)}{4\pi b(\xi)}\:.
    \label{eq:Temperature-From-a-b-functions}
\end{align}
The scalar $\phi(\xi)$ enters the temperature indirectly since the constraint equation relates $b(\xi)$ to $a(\xi)$ and $\phi(\xi)$. In the limit of no back-reaction, $a(0) = \alpha^{-1/4}$, $b(0)= \alpha^{-1/2}/4$, so that we get the familiar result
\begin{align}
    T = \frac{\rho_\uv}{\pi}\left(\frac{\alpha}{1+\alpha}\right)^{1/4} = \frac{\rho_h}{\pi}\:,
    \label{eq:Temperature-noBR}
\end{align}
where in the last line we used $\rho_h = \rho_\uv (\alpha/(1+\alpha))^{1/4}$. This shows that smaller $\alpha$ corresponds to smaller $T$. As we will see, this is not true once there is significant back-reaction.

\subsection{Free Energy}
\label{subsec:FreeEnergy}
Ref.~\cite{Creminelli:2001th} computed the free energy by evaluating the partition function in the Euclidean signature at the saddle corresponding to the solution of the equations of motion. However, this approach is both subtle and tedious due to the UV-sensitive counter-terms that have to be removed by holographic renormalization, which in the present case would require to be removed numerically leading to a finite part which also would be evaluated numerically. Here we take an alternate approach that circumvents these issues. Using $F = -(1/\beta)\log Z$ and $S = (1-\beta\partial_\beta)\log Z$, we get $S = \beta^2\partial_\beta F = -\partial_T F$. Note that both the free energy $F$ and the entropy $S$ are extensive, so we also have $s = -\partial_T f$ where $s = S/\text{Vol}_3, f = F/\text{Vol}_3$. The relation between $s$ and $f$ can be inverted to give
\begin{align}
    f - f_0 = -\int \dd T \,s(T)  = -\int \dd \alpha \frac{\dd T(\alpha)}{\dd \alpha}\,s(\alpha) \:,
\end{align}
where $f_0$ is an overall constant, and we have changed variables in the last line. The advantage of this formula is that we can use the entropy calculated in the previous subsection (eq.~\eqref{eq:entropy-BH}), and the only non-trivial step is to calculate $T(\alpha)$.

As a check, in the limit of no back-reaction, using $T/\rho_\uv = \alpha^{1/2}/(1+\alpha)^{1/4}$ (from eq.~\eqref{eq:Temperature-From-a-b-functions}) and the expression for entropy in the limit of no back-reaction (from eq.~\eqref{eq:entropy-BH}), we get $f = -2\pi^4 M_5^3 T^4$, which is the expected answer, computed in a different manner in ref.~\cite{Creminelli:2001th}. We see that using the relation $s = -\partial_T f$ circumvents the need for pesky counter-term subtraction and rescaling of temperatures at the UV cutoff.

\subsection{Speed of Sound and the Specific Heat}
\label{subsec:SOS-SH}
Given the entropy $s$ and the temperature $T$, one can also obtain the speed of sound $c_s$ and the specific heat $C_V$ with the following identities:
\begin{align}
    c_s^2 &= \frac{\dd \log T}{\dd \log s}\, , \:\:
    C_V = \frac{\dd s}{\dd \log T} = \frac{s}{c_s^2}\:.
    \label{eq:speed-of-sound-exact-def}
\end{align}
In the limit of no back-reaction, $s = 8\pi^4 M_5^3 T^3$ so that $c_s^2 = 1/3$ and $C_V = 24\pi^4 M_5^3 T^3$. As we will show in the later sections, $c_s^2$ decreases and crosses zero from above as the horizon is pushed in the IR, when the effect of back-reaction is important. The entropy is finite as $c_s^2$ crosses zero, and $C_V$ has an infinite discontinuity across this point. Crucially, this is suggestive of a second-order phase transition, which is a point we will return to.

\section{Numerical Results}
\label{sec:results}
In this section we present the numerically obtained scalar profiles and the temperature as a function of $\alpha$, for several choices of parameters.  For these parameters, we find the free energy and entropy as a function of temperature. We show a significant effect from the back-reaction for some choices of parameters,  which results in a minimum temperature as a function of $\alpha$. When this happens, the entropy and free energy as a function of temperature have a turn-around at the minimum temperature and are multiple-valued for higher temperatures. 

We first present our choice of parameters. For all the choices, we always choose $\phi_\uv = 1$. We choose a few values of $\epsilon$ and $\phi_c$ to show the effect of back-reaction. We work with values of $\alpha$ in the range $(10^{-4}-10^{-7})$; lower values of $\alpha$ lead to numerical issues. 

Given that $\alpha\approx(\rho_h/\rho_\uv)^4$, one might worry that we have not generated a large hierarchy between the UV and the IR. Here we have implicitly assumed that a hierarchy has already been generated, between some ultra UV (UUV) scale (where $\phi = \phi_\text{uuv} \ll 1$) and what we call UV (where $\phi_\uv = 1$). This works for our choice of a negative $\epsilon$, for which $\phi$ grows monotonically towards the IR. Recall that we have normalized the coordinate $\xi$ such that $\xi = 1$ corresponds to the UV, and therefore UUV corresponds to $\xi = \xi_\text{uuv} \gg 1$. Once we have a solution for the scalars in the range $0 \leq \xi \leq 1$, we can solve the differential equation from $\xi = 1$ to larger values of $\xi$, to get the required value of $\phi$ in the UUV, given a hierarchy, and vice versa. This approach is to ensure that we obtain numerically robust results, and focus directly on the IR where strong modifications occur due to a large back-reaction, while the hierarchy is generated perturbatively in the UUV region where the back-reaction is small. 

Depending on the magnitude of $\epsilon$, starting with $\phi = \phi_\uv = 1$ at $\xi = 1$, $\phi$ grows to some value $\phi_h$ as the horizon $\xi \to 0$ is approached. Depending on how far the horizon is from the UV, dictated by the parameter $\alpha$, $\phi_h$ can be close enough to $\phi_c$ to get a substantial modification to the isolated gravitational solution.  We expect this effect to show up at smaller values of $\alpha$, when $\epsilon$ is reduced in magnitude, or when $\phi_c$ is increased. 

We choose the following parameters:  
\begin{align}
    \textbf{A:} & \qquad \epsilon = -1/50, \:\phi_c = 3/2\:. 
    \nonumber \\
    \textbf{B:} & \qquad \epsilon = -1/60, \:\phi_c = 3/2\:.
    \nonumber \\
    \textbf{C:} & \qquad \epsilon = -1/100, \:\phi_c = 3/2\:.
    \nonumber \\    
    \textbf{D:} & \qquad \epsilon = -1/60, \:\phi_c = 2\:.
    \label{eq:params}
\end{align}
In all these, we choose $\phi_\uv = 1$ and $\xi_\ir/\alpha = 10^{-2}$.

\subsection{Scalar Profiles}
\begin{figure}[h!]
    \centering
    \includegraphics[width=0.45\textwidth]{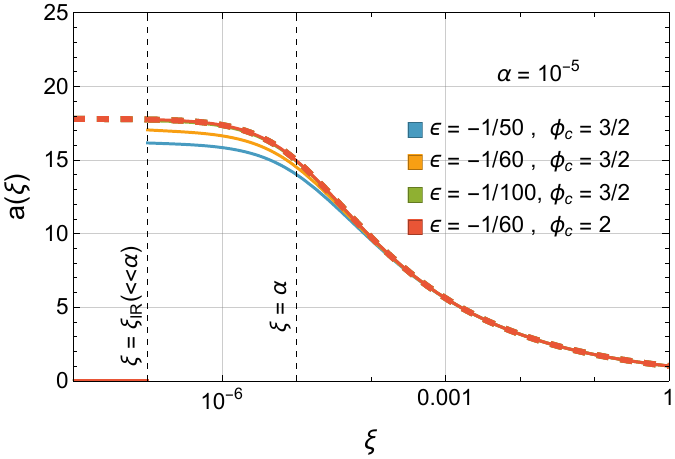}
    \includegraphics[width=0.45\textwidth]{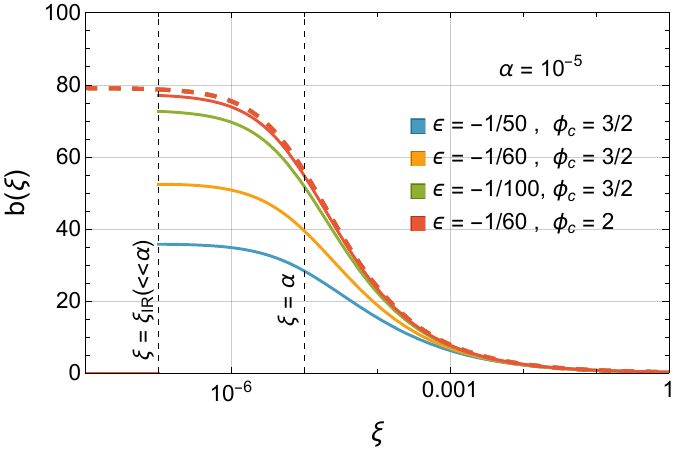}
    \includegraphics[width=0.45\textwidth]{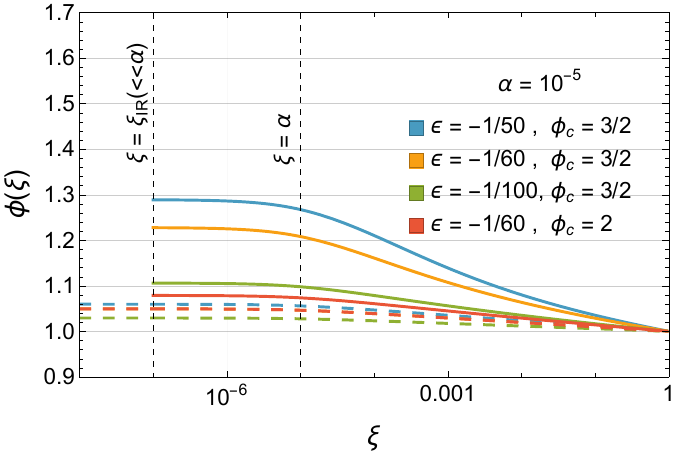}
    \caption{\small{Scalar profiles for the parameters in eq.~\eqref{eq:params}, for $\alpha = 10^{-5}$. In each plot, the horizon is at $\xi = 0$ (left) and the UV is at $\xi = 1$ (right).  Also shown are the reference profiles (with $\phi_c \to \infty$) in dashed color. For $a(\xi), b(\xi)$, the reference profiles (in dashed) are the same for each color because $\phi$ decouples from $a, b$ in the limit of small back-reaction.}}
    \label{fig:scalar-profiles-fixedAlpha}
\end{figure}

We first look at the scalar profiles $\phi(\xi), a(\xi)$ and $b(\xi)$ (which is fixed in terms of $\phi(\xi)$ and $a(\xi)$ by the constraint, eq.~\eqref{eq:b-eqn-constraint}) for the parameter choices in eq.~\eqref{eq:params}. We first show numerical solutions for fixed $\alpha$, i.e. the horizon is at a fixed coordinate distance from the UV. Fig.~\ref{fig:scalar-profiles-fixedAlpha} shows the profiles for $\alpha = 10^{-5}$ for the parameter choices $\textbf{A,B,C,D}$. The solid lines in color show the profiles while the dashed lines in color show the reference behavior when there is no back-reaction (i.e. $\phi_c\to\infty$ limit). Also shown in vertical dashed are the locations $\xi = \alpha$ from where the effect of a horizon starts to show up (the profiles stop growing) and $\xi = \xi_\ir \ll \alpha$, which is the IR cutoff we work with. 
Several trends can be noted from the profiles:
\begin{itemize}
    \item The growth of $\phi$ in the IR is largest for the largest $\left|\epsilon\right|$ and smallest $\phi_c$ (blue curve in lower panel of Fig.~\ref{fig:scalar-profiles-fixedAlpha}), and drops as $\left|\epsilon\right|$ is reduced or $\phi_c$ is increased. Also shown are the reference profiles (with $\phi_c \to \infty$) in dashed. 
    \item Both $a(\xi)$ and $b(\xi)$ reduce due to the back-reaction, as seen by e.g. the blue line being the lowest in the top left and top right panels of Fig.~\ref{fig:scalar-profiles-fixedAlpha}.
    \item The reference plots in dashed lines are the same for all the colors: without back-reaction, $a,b$ are decoupled from $\phi$.
\end{itemize}

\begin{figure}[h!]
    \centering
    \includegraphics[width=0.45\textwidth]{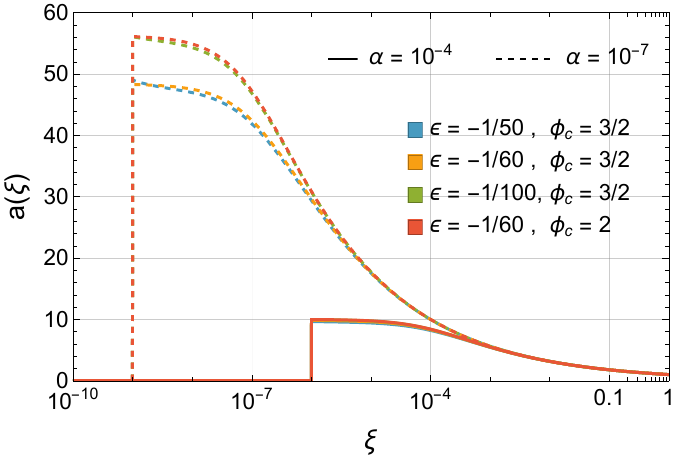}
    \includegraphics[width=0.45\textwidth]{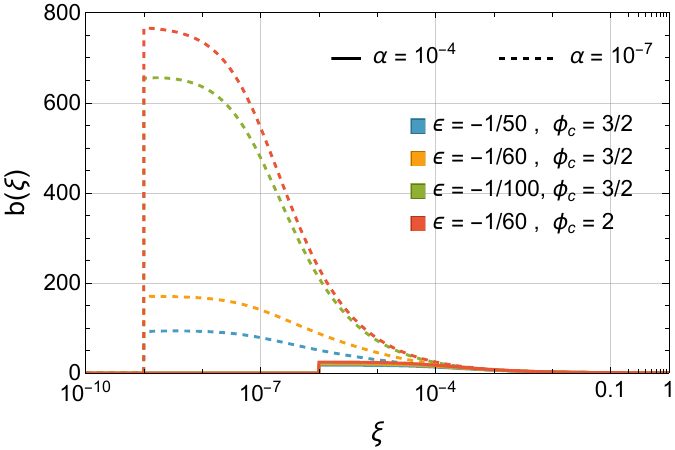}
    \includegraphics[width=0.45\textwidth]{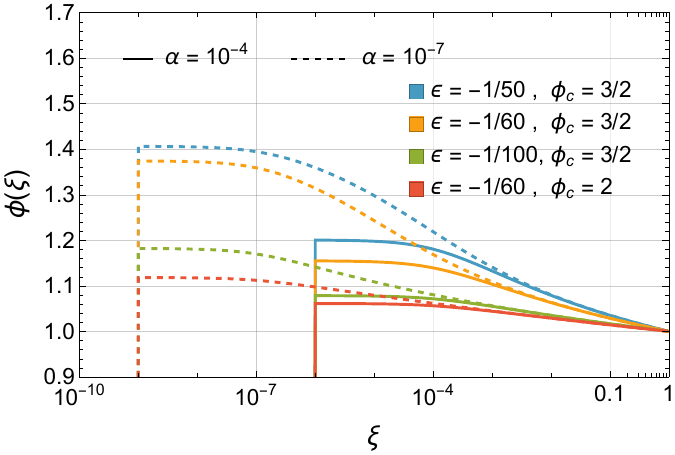}
    \caption{\small{Scalar profiles for the parameters in eq.~\eqref{eq:params}, for $\alpha = 10^{-4}$ (solid) and $10^{-7}$ (dotted). In each plot, the horizon is at $\xi = 0$ (left) and the UV is at $\xi = 1$ (right).}}
    \label{fig:scalar-profiles-varyAlpha}
\end{figure}

We next compare the profiles for a few values of $\alpha$, with other parameters fixed. Fig.~\ref{fig:scalar-profiles-varyAlpha} shows the profiles for $\alpha = 10^{-4}$ (in solid colors) and $\alpha = 10^{-7}$ (in dotted colors). 
Once again several trends can be noted from the profiles:
\begin{itemize}
    \item For smaller $\alpha$, all the profiles grow more in the IR, and are eventually saturated as $\xi$ gets smaller than the corresponding $\alpha$. 
    \item The color order for $\phi$ (i.e. blue growing the most and red the least) persists as $\alpha$ is made smaller. This order is reversed for $a(\xi), b(\xi)$, and that too persists.
\end{itemize}

In summary, the profiles have the behavior expected on general grounds. We now proceed to calculate the derived thermodynamic quantities from these profiles in the next subsections. 

\subsection{Temperature}
\begin{figure}[h!]
    \centering
    \includegraphics[width=0.99\textwidth]{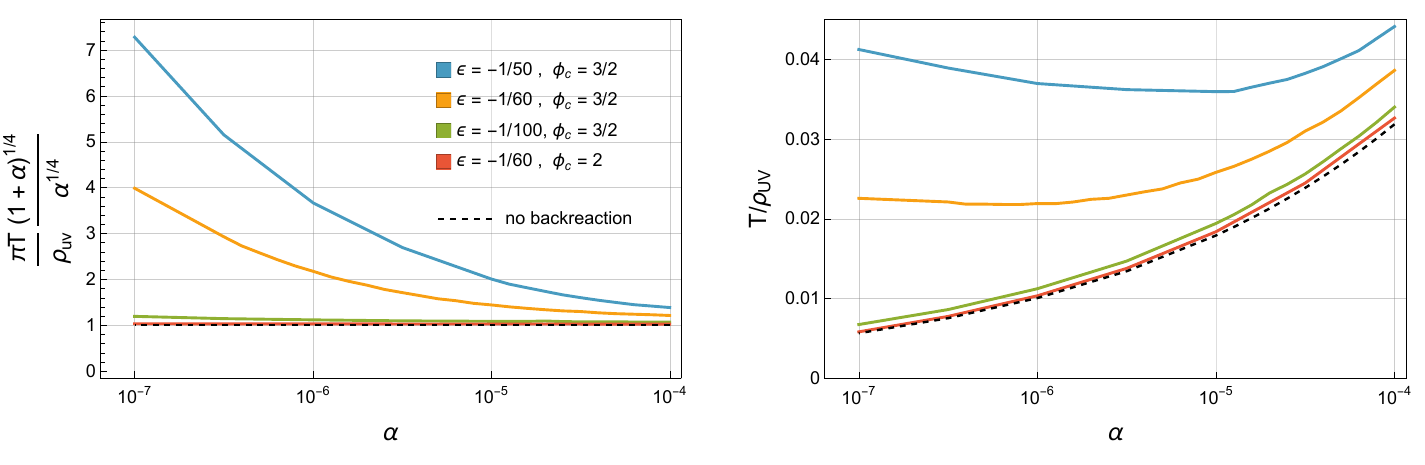}
    \caption{\small{Temperature as a function of $\alpha$: on left is $(\pi T(\alpha)/\rho_\uv)(1+1/\alpha)^{1/4}$ (which is unity in the limit of no back-reaction), while on the right $T(\alpha)/\rho_\uv$ is shown directly. A minimum temperature is observed for the blue and orange curves.}}
    \label{fig:temperature-vs-alpha}
\end{figure}

We would like to calculate thermodynamic quantities of interest in the canonical ensemble, i.e. as a function of temperature. Having obtained the profiles, it is straightforward to obtain the temperature of the deconfined phase, as given by eq.~\eqref{eq:Temperature-From-a-b-functions}. In fig.~\ref{fig:temperature-vs-alpha} we show the temperature as $\alpha$ is varied. We work with the ratio $T(\alpha)/\rho_\uv$. The left panel shows $(\pi T(\alpha)/\rho_\uv)(1+1/\alpha)^{1/4}$ as a function of $\alpha$. In the limit of no back-reaction $(\pi T(\alpha)/\rho_\uv)(1+1/\alpha)^{1/4}$ equals unity, as can be seen by plugging no back-reaction values of $a(0), b(0)$ in eq.~\eqref{eq:Temperature-From-a-b-functions}. The right panel shows $T(\alpha)/\rho_\uv$ directly. In both, the no back-reaction case is shown by the black dotted line. 
One can clearly see a modification in temperature from the back-reaction. In the left panel of fig.~\ref{fig:temperature-vs-alpha}, compared to the no back-reaction value, the rescaled $T$ increases as $\alpha$ gets smaller. The order of the colors (i.e. the blue having the highest temperature and the red having the lowest temperature at smaller values of $\alpha$) is in agreement with the expectation---the parameter with a larger back-reaction has a larger increase in $T$. In the right panel of fig.~\ref{fig:temperature-vs-alpha}, one can see that the behavior of $T(\alpha)/\rho_\uv$ as a function of $\alpha$ is very different in the presence of significant back-reaction: for the parameters $\textbf{A, B}$ the temperature initially reduces but then starts to increase as $\alpha$ reduces. As elaborated in sec.~\ref{subsec:Entropy}, smaller $\alpha$ means smaller entropy density of the deconfined phase. Since temperature is no longer monotonically dependent on $\alpha$ when there is back-reaction, the relation between temperature and entropy is modified, which we show later. The very interesting result here is a minimum temperature.  

\subsection{Proper Distance of Horizon from UV}
\label{subsec:proper-distance-of-horizon}
To understand the effect of the back-reaction on the geometry better, it is instructive to look at the coordinate invariant proper distance of the horizon from the UV. Starting with the metric parametrized by eq.~\eqref{eq:AdSS-metric-ansatz}, the proper distance between the UV (at $\xi = 1$) and the horizon (at $\xi = 0$) is given as
\begin{align}
    d_H &= \int_0^1 \frac{\dd \xi}{\sqrt{\xi}}\,b(\xi)\:.
    \label{eq:proper-distance}
\end{align}
In the limit of no back-reaction, this reduces to
\begin{align}
    d_H \:\: \underset{\text{no b.r.}}{=} \:\:
    \int_0^1 \frac{\dd \xi}{4\sqrt{\xi}\sqrt{\xi + \alpha}}
    = \frac12 \tanh^{-1}\left(\frac{1}{\sqrt{1+\alpha}}\right) 
    \underset{\alpha \ll 1}{\approx} \frac{2}{\sqrt{\alpha}} + \mathcal{O}(\alpha^{-3/2}) \:,
    \label{eq:HorizonProperDistance}
\end{align}
and is a monotonically decreasing function of $\alpha$---smaller values of $\alpha$ mean the horizon is farther  from the UV.

\begin{figure}[h!]
    \centering
    \includegraphics[width=0.48\textwidth]{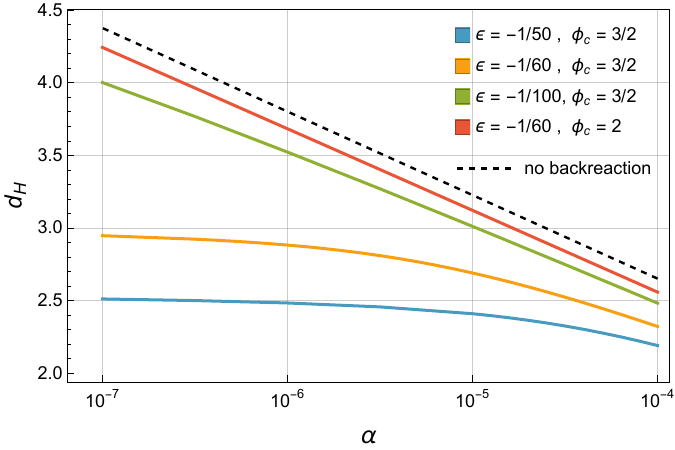}
    \includegraphics[width=0.48\textwidth]{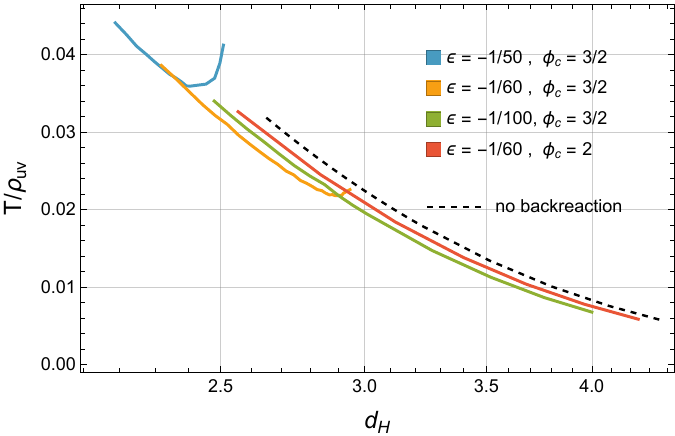}
    \caption{\small{Left: proper distance (see eq.~\eqref{eq:proper-distance}) as a function of $\alpha$. For parameters with significant back-reaction, the proper distance starts to saturate at lower $\alpha$. Right: Temperature as a function of proper distance. A sharp rise beyond the point with minimum value is noticeable, sharpest for the case with the strongest back-reaction.}}
    \label{fig:Temperature-vs-ProperDistance-vs-alpha}
\end{figure}

We can calculate the proper distance numerically for the parameter choices in eq.~\eqref{eq:params}. Since we are working with an IR cutoff $\xi_\ir \neq 0$, we change the lower limit of the integration in eq.~\eqref{eq:proper-distance} to $\xi = \xi_\ir$. The left panel of fig.~\ref{fig:Temperature-vs-ProperDistance-vs-alpha} shows the proper distance $d_H$ as a function of $\alpha$ for the parameter choices. The salient feature is that due to back-reaction, the proper distance starts to saturate as $\alpha$ is lowered. This effect is stronger for parameters with the largest back-reaction. 

The right panel of fig.~\ref{fig:Temperature-vs-ProperDistance-vs-alpha} shows temperature as a function of proper distance. The sharp turn-around in temperature is very prominent for the blue and orange curves, which have the most significant back-reaction. The sharper rise in temperature beyond its minimum value (as compared to fig.~\ref{fig:temperature-vs-alpha}) is because the proper distance changes very little beyond a certain point, and this shows up as a sharp rise in temperature as a function of the proper distance. 
Note that in the small back-reaction case, when eq.~\eqref{eq:HorizonProperDistance} holds, a smaller $\alpha$ translates to a larger $d_H$, and the temperature changes monotonically as a function of $\alpha$ (see eq.~\eqref{eq:Temperature-noBR}). However as fig.~\ref{fig:Temperature-vs-ProperDistance-vs-alpha} shows, for large back-reaction, making $\alpha$ smaller changes $d_H$ only by a small amount compared to the no back-reaction case.

\subsection{Entropy vs Temperature}
\begin{figure}[h!]
    \centering
    \includegraphics[width=0.99\textwidth]{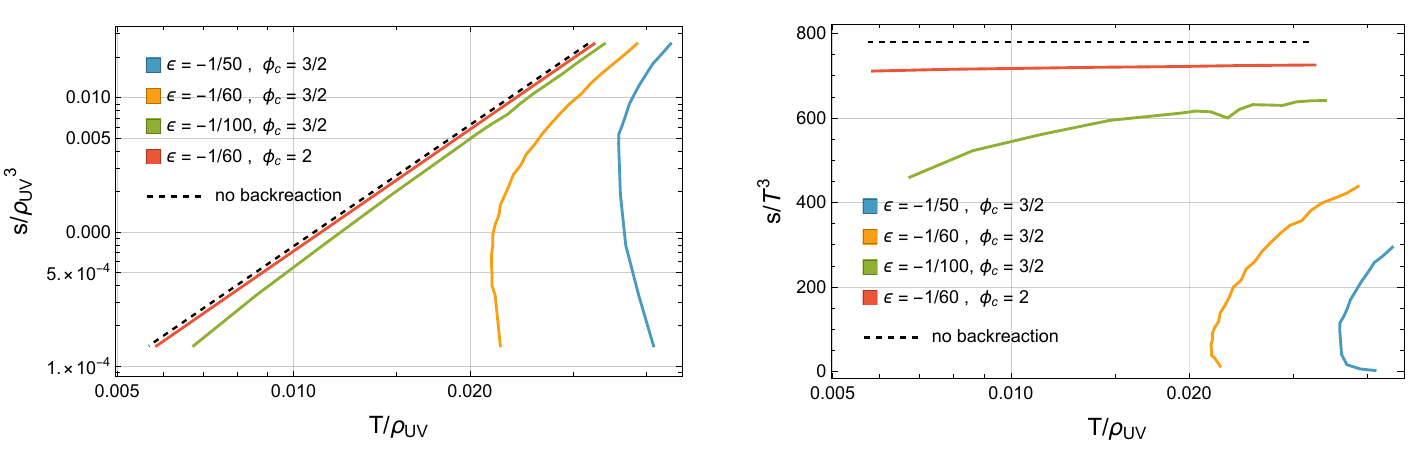}
    \caption{\small{Left: Entropy as a function of temperature. Cases with strong back-reaction (blue and orange) show a turn-around, suggesting the phase becomes unstable. Right: $s/T^3$, a measure of degrees of freedom, as a function of temperature. The drop in the degrees of freedom is strongest for the cases with the most back-reaction. $M_5$ has been set to 1 in all these plots.}}
    \label{fig:entropy-DOF-withBR}
\end{figure}

As we saw in sec.~\ref{subsec:Entropy}, the expression for entropy as a function of $\alpha$ is the same with or without back-reaction. The relation between the temperature and $\alpha$ is however different in the presence of back-reaction, and this changes the $s-T$ phase diagram. The left panel of fig.~\ref{fig:entropy-DOF-withBR} shows the entropy as a function of temperature for the parameter choices we consider. The turn-around in entropy around a minimum temperature is evident for the blue and orange curves where back-reaction is significant. This suggests that once the temperature crosses the minimum temperature, the hot phase is not of the highest entropy, and is thermodynamically unstable.

This also suggests the standard first-order phase transition story is modified when there is a strong back-reaction. For the case of no back-reaction, the entropy has a standard behavior and is shown in black dashed in the left panel of fig.~\ref{fig:entropy-DOF-withBR}. In the right panel of fig.~\ref{fig:entropy-DOF-withBR}, we plot $s/T^3$ which is a measure of degrees of freedom, as a function of temperature. For the cases with the strongest back-reaction, the degrees of freedom effectively approaches zero much faster than the no back-reaction or small back-reaction cases (see also ref.~\cite{Megias:2018sxv} which studied the reduction in degrees of freedom from back-reaction). The odd upturn is again related to the existence of a minimum temperature. 

\subsection{Free Energy vs Temperature}
\label{sec:NumericalResults-FE}
Finally, we can calculate the free energy using the steps outlined in~\ref{subsec:FreeEnergy}. Since $s(T)$ is a multi-valued function of temperature, we have to be careful in evaluating the integral. We change variables from $T$ to $s$:
\begin{align}
    f(T) - f_0 = -\int_{T_0}^T \dd T\,s(T) = -\int_{s_0}^s \dd s \frac{\dd T}{\dd s} s = -\int_{s_0}^s \dd s \, T \frac{\dd \log T}{\dd \log s} \:.
    \label{eq:FE-from-Entropy}
\end{align}
Since $T(s)$ is a single-valued function, there is no complication. The integral will give $f(s)$ and since we know $s(T)$, we know $f(T)$ parametrically. To obtain $T(s)$, we work with a fitted function rather than the discrete numerical data. We use the following functional form:
\begin{align}
    \log \, T(s) &= a_0 + a_1\,\log s + a_2\, \left(\log s\right)^2 + a_3\,\left(\log s\right)^3\: .
    \label{eq:FunctionToFit-T-entropy}
\end{align}
\begin{figure}[h!]
    \centering
    \includegraphics[width=0.99\textwidth]{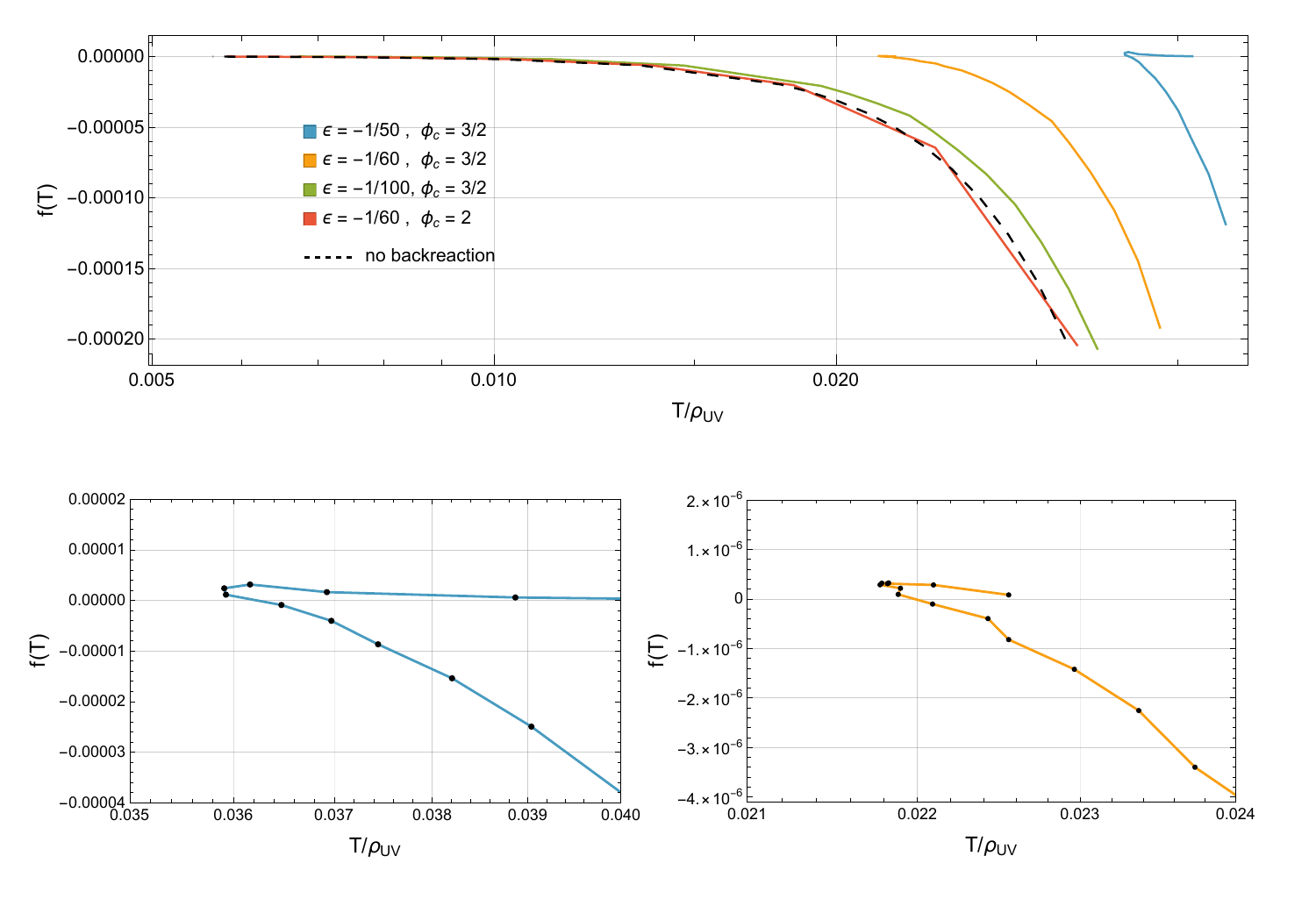}
    \caption{\small{Free energy as a function of temperature. The top panel shows the behavior for all four benchmark points, with the no back-reaction case also shown. The lower panels show the zoomed-in behavior of the two cases with the most significant back-reactions, the one in blue and orange, in detail near their turn around point (the numerical data points are shown in black dots). $M_5$ has been set to 1 in all these plots.}}
    \label{fig:FE-vs-Temperature}
\end{figure}
Figure~\ref{fig:FE-vs-Temperature} shows free energy as a function of temperature for the parameter choices in eq.~\eqref{eq:params}. The top panel shows the free energy for the four parameter choices, also showing the no back-reaction expectation in black dashed. The free energy has a turn around for the cases with significant back-reaction. In the lower two panels of fig.~\ref{fig:FE-vs-Temperature}, we zoom in near the turn-around point to show the effect more clearly---the colors are in one to one correspondence to the top panel.

As we will elaborate in the discussion section (sec.~\ref{sec:discussion}), the turn-around and a shark-fin like feature in the $f-T$ diagram is reminiscent of the shape in the analysis of ref.~\cite{Buchel:2021yay}---this was the task we set out for, to model the features seen in the deformed conifold geometry using a scalar with suitable potential.

\section{Analytical Estimates}
\label{sec:analyticalEstimates}
In the absence of a closed-form solution we chose to proceed numerically as was presented in the previous section. However, we can give some analytical estimates to provide non-trivial checks of our results.  The steps in this section use results from ref.~\cite{Gubser:2008ny}.

For analytical results, it is more convenient to work with a different parameterization of the metric. Consider the metric
\begin{align}
    \dd s^2 &= e^{2 A(r)}\left(-f(r)\,\dd t^2 + \dd x^2\right) + e^{2B(r)}\frac{\dd r^2}{f(r)}\:, \qquad r_\uv \leq r \leq r_h\:,
    \label{eq:metric-r-coordinates}
\end{align}
where the UV brane is at $r = r_\uv$ and the horizon is at $r = r_h$. There are three scalar functions $A(r), B(r)$ and $f(r)$ that parameterize the spacetime. The scalar $\sigma$ also has a profile in the radial direction, $\sigma(r)$. The gauge redundancy in choosing the radial direction $r$ can be fixed by using $\sigma$ itself as the radial direction. The metric then becomes
\begin{align}
    \dd s^2 &= e^{2 A(\sigma)}\left(-f(\sigma)\,\dd t^2 + \dd x^2\right) + e^{2B(\sigma)}\frac{\dd \sigma^2}{f(\sigma)}\:, \qquad \sigma_\uv \leq \sigma \leq \sigma_h\:.
    \label{eq:metric-sigma-coordinates}
\end{align}
\subsection{Temperature and Entropy}
\label{subsec:TemperatureAndEntropy}
In the rest of the section, we work with the metric in eq.~\eqref{eq:metric-sigma-coordinates}. The horizon location $\sigma = \sigma_h$ is defined by 
\begin{align}
    f(\sigma = \sigma_h) = 0\:.
\end{align}
The temperature can be calculated in these coordinates as before, by requiring smoothness of the Euclidean metric with time compactified. Expanding the Euclidean version of the metric in eq.~\eqref{eq:metric-sigma-coordinates} around $\sigma = \sigma_h$ and using $f(\sigma_h) = 0$, we get (in the $(t_E, \sigma)$ space)
\begin{align}
    \dd s^2 \underset{\sigma\to \sigma_h}{ = } e^{2A(\sigma_h)}\,f'(\sigma_h)\left(\sigma - \sigma_h\right)\dd t_E^2 + e^{2B(\sigma_h)}\frac{\dd \sigma^2}{f'(\sigma_h)(\sigma-\sigma_h)}\:.
    \label{eq:metric-sigma-coordinates-nearHorizon}
\end{align}
Changing coordinates from $\sigma$ to $r = 2\sqrt{(\sigma - \sigma_h)/f'(\sigma_h)} e^{B(\sigma_h)}$ ($r = 0$ at $\sigma = \sigma_h$), the metric in eq.~\eqref{eq:metric-sigma-coordinates-nearHorizon} becomes
\begin{align}
    \dd s^2 = \left(\frac12 e^{A(\sigma_h)-B(\sigma_h)} f'(\sigma_h)\right)^2 r^2\,\dd t_E^2 + \dd r^2\:.
\end{align}
Given that $t_E$ is periodic with period $1/T$, the above metric is smooth (i.e. reduces to flat space in polar coordinates) when
\begin{align}
    T &= e^{A(\sigma_h) - B(\sigma_h)}\,\frac{\left|f'(\sigma_h)\right|}{4\pi} \:.
    \label{eq:Temperature-sigma-coordinates}
\end{align}

The entropy (per unit 3-volume) $s$ can be calculated as before by using the Bekenstein-Hawking formula, and is given by
\begin{align}
    s = \frac{S}{\text{Vol}_3} = \frac{\mathcal{A}_\text{horizon}}{4G_N}\frac{1}{\text{Vol}_3} = 8 \pi M_5^3 e^{3 A(\sigma_h)}\:,
    \label{eq:Entropy-sigma-coordinates}
\end{align}
where we have used $1/G_N = 32\pi^2 M_5^3$.

Leaving the details to app.~\ref{app:AnalyticalResults}, the equations of motion for $f(\sigma), A(\sigma), B(\sigma)$ can be manipulated to obtain 
\begin{align}
    \frac{\dd \log T}{\dd \sigma_h} & \approx -\frac{V(\sigma_h)}{V'(\sigma_h)}\left(\frac13 - \frac12 \left(\frac{V'(\sigma_h)}{V(\sigma_h)}\right)^2\right)\:,
    \label{eq:Temperature-vs-horizon-approx}
    \\
    \frac{\dd \log s}{\dd \sigma_h} &\approx - \frac{V(\sigma_h)}{V'(\sigma_h)}\:.
    \label{eq:Entropy-vs-horizon-approx}
\end{align}
The approximate sign in the above comes from dropping $\partial_{\sigma_h}$ terms in the so-called ``adiabatic approximation'' (see app.~\ref{app:AnalyticalResults} and ref.~\cite{Gubser:2008ny, Zollner:2018uep} for details.)

Using the approximate expressions~\eqref{eq:Temperature-vs-horizon-approx} and~\eqref{eq:Entropy-vs-horizon-approx}, we can further obtain the speed of sound squared, the specific heat, and the effective degrees of freedom $s/T^3$ to be:
\begin{align}
    c_s^2 
    &\approx \frac13 - \frac12\left(\frac{\partial_\sigma V(\sigma_h)}{V(\sigma_h)}\right)^2\:.
    \label{eq:speed-of-sound-approx}
    \\
    C_V 
    &\approx 8 \pi M_5^3 e^{3A(\sigma_h)}\left(\frac13 - \frac12\left(\frac{\partial_\sigma V(\sigma_h)}{V(\sigma_h)}\right)^2\right)^{-1}
    \label{eq:specific-heat-approx}
    \\
    s/T^3 &\propto \left|V(\sigma_h)\right|^{-3/2}\:.
    \label{eq:dof-approximate}
\end{align}
The details of the manipulations leading to these results is again relegated to app.~\ref{app:AnalyticalResults}. There are several interesting features one can see from these equations. Eq.~\eqref{eq:Temperature-vs-horizon-approx} shows that $\dd \log T/\dd \sigma_h$ changes sign depending on the value of $V'/V$ at the horizon. If $V'/V$ at the horizon is small, $\dd \log T / \dd \sigma_h$ is negative, i.e. the temperature decreases as $\sigma_h$ increases. Since the scalar $\sigma$ grows in the IR, this means the temperature decreases as the horizon moves farther into the IR. This is the usual result for an AdSS spacetime, without a significant back-reaction from a scalar. 

However when $V'/V$ is larger than $\sqrt{2/3}$, $\dd \log T /\dd \sigma_h$ becomes positive, and the temperature starts to increase as $\sigma_h$ increases. The turn-around point is the value of $\sigma_h$ where the temperature attains a minimum, at $V'/V = \sqrt{2/3}$. Eq.~\eqref{eq:speed-of-sound-approx} shows that the speed of sound goes to zero from above at $V'/V = \sqrt{2/3}$, and is correlated with the minimum temperature. These two effects, the minimum temperature and a vanishing speed of sound are also correlated with the specific heat approaching positive infinity. Finally, eq.~\eqref{eq:dof-approximate} shows that as the potential grows (in magnitude) in the IR, the effective degrees of freedom reduce, as expected from a holographic field theory point of view. These statements are however only as good as the adiabatic approximation is.\footnote{Note that for a purely exponential potential $V = V_0 e^{\gamma \sigma}$, with a specific relation between $V_0$ and $\gamma$, referred in the literature as the Chamblin-Reall (CR) solution~\cite{Chamblin:1999ya}, eq.~\eqref{eq:speed-of-sound-approx} is exactly correct. The CR solution is obtained by starting with a $d+1$ dimensional pure AdSS solution and compactifying $d-4$ dimensions on a torus---the resulting 5D action is simply Einstein gravity plus a scalar with the exponential potential $V_0 e^{\gamma\sigma}$, with its coefficient and exponent fixed in terms of the $d+1$ dimensional AdS radius, and the quantity $d$.}

These steps also make clear the utility of using the scalar $\sigma$ as the radial coordinate (eq.~\eqref{eq:metric-sigma-coordinates}) for the analytical steps in this section, as opposed to the $\xi$ coordinates used earlier (eq.~\eqref{eq:AdSS-metric-ansatz}). The coordinate systems trade $(\xi, \alpha)$ with $(\sigma, \sigma_h)$. In the $\xi$ coordinate system, the horizon is always at $\xi = 0$ and allows obtaining a series solution near the horizon easily (as done in sec.~\ref{subsec:boundary-conditions} to obtain the IR boundary conditions). In the $\sigma$ coordinate system the horizon is at $\sigma = \sigma_h$ and $\sigma_h$ can vary. The $\sigma$ coordinate system allows separating $\partial_\sigma$ and $\partial_{\sigma_h}$ effects, to obtain the results in the adiabatic approximation, which are obscured in the $\xi$ coordinate system. 

The $\sigma$ coordinate system can also be used to obtain numerical results (as we show in app.~\ref{app:numericalMethod-Alternate}), but makes several effects less manifest.  As we saw in section~\ref{sec:results}, the growth of the scalar slows down near the horizon, which would be less clear if the scalar is itself used as the coordinate. Further, since the scalar decouples from the dynamics in the no back-reaction case, this choice is not well suited to compare the results obtained in literature where back-reaction is ignored. Finally, in the scalar coordinates, the horizon area is not fixed just in terms of parameters, but one rather has to solve for it numerically (see eq.~\eqref{eq:Entropy-sigma-coordinates}). For these reasons, we have used the $\xi$ coordinates in the main text.

\subsection{Analytical Understanding of Free Energy}
\begin{figure}[h!]
    \centering
    \includegraphics[width=0.7\textwidth]{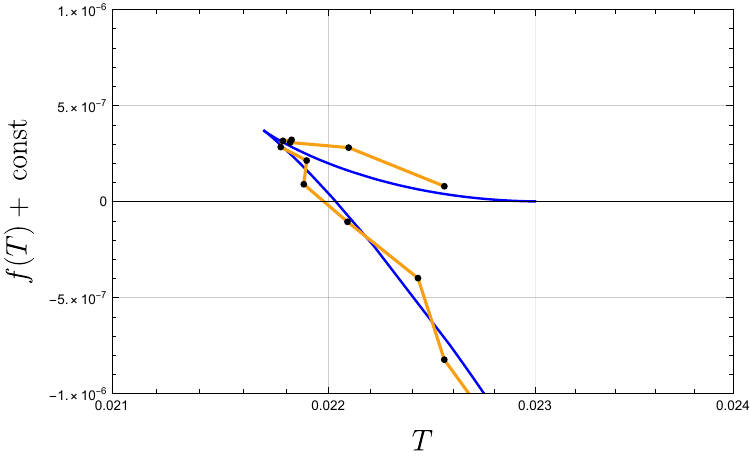}
    \caption{\small{Free energy as a function of temperature: blue line is calculated using eq.~\eqref{eq:T-vs-s-quadratic}, while the black dots (joined by orange line) are numerical results for parameter \textbf{B} in eq.~\eqref{eq:params}. The curves match well near $T_\text{min}$ where eq.~\eqref{eq:T-vs-s-quadratic} is a good approximation.}}
    \label{fig:FE-cartoon-1}
\end{figure}

Using the analytical steps, we have argued that the temperature has a minimum value, as seen by eq.~\eqref{eq:Temperature-vs-horizon-approx}. Using this, we can also understand the sharp turn-around in the free energy as a function of temperature. Recall that the entropy as a function of $\alpha$ (eq.~\eqref{eq:entropy-BH}) is monotonic, while the temperature has a minimum at $T= T_\text{min}$. The function $s(T)$ is not single-valued, but the inverse function $T(s)$ is. In the vicinity of $T = T_\text{min}$, the function $T(s)$ can be approximated as a quadratic function: 
\begin{align}
    T(s) \approx T_\text{min} + \lambda (s - s_0)^2\:, 
\end{align}
where $s_0$ is the value of entropy at $T = T_\text{min}$. This gives two branches of solutions (referred as $\pm$):
\begin{align}
    s_\pm(T) = s_0 \pm \sqrt{\frac{T-T_\text{min}}{\lambda}}\:.
    \label{eq:T-vs-s-quadratic}
\end{align}
We can calculate the free energy using the thermodynamic relation $s = -\partial_T f$: 
\begin{align}
    f(T) + \text{const} = -\int \dd T \, s(T) = -\int \dd s \, \frac{\dd T}{\dd s} \, s\:.  
\end{align}
Note that since $s_+ > s_-$ for $T > T_\text{min}$, the slope of the two branches for the free energy $f_\pm$ are different at $ T > T_\text{min}$.
Evaluating $f$ for the $\pm$ branches, we get
\begin{align}
    f_\pm (T) = -\frac{2\lambda}{3}\left(s_0 \pm \sqrt{\frac{T - T_\text{min}}{\lambda}}\right)^3
    \left(1 - \frac32 \frac{s_0}{s_0 \pm \sqrt{\frac{T - T_\text{min}}{\lambda}}}\right)\:.
\end{align}
Fig.~\ref{fig:FE-cartoon-1} compares the approximate expression for $f_\pm(T)$ with numerical results obtained earlier (in $\xi$ coordinate system, for parameter \textbf{B}): blue lines show $f_\pm(T)$ for $\lambda = 2000, s_0 = 0.0008, T_\text{min} = 0.02$, while black dots (joined by orange line) are numerical results for parameter \textbf{B}. The numerical value of $\lambda, s_0, T_\text{min}$ are obtained by fitting a quadratic to the $(T,s)$ data.  

\subsection{Analytical Understanding of Speed of Sound}
We can use the discussions in sec.~\ref{subsec:TemperatureAndEntropy} to understand the numerical results obtained earlier. Using eq.~\eqref{eq:speed-of-sound-approx}, the speed of sound squared is given explicitly as (using $V(\sigma)$ from eq.~\eqref{eq:potential_sigma})
\begin{align}
    c_s^2 = \frac13 - \frac{2}{9 \sigma_c^2}  \left(\frac{5 - 7\, e^{-\sigma_h/\sigma_c}+ 2\, e^{-2\,\sigma_h/\sigma_c}}{1-2\,e^{-\sigma_h/\sigma_c} +e^{-2\,\sigma_h/\sigma_c} + (\Lambda/\widetilde{\epsilon}\sigma_c^2)\, e^{-(10/3)\,\sigma_h/\sigma_c}}\right)^2\:,
    \label{eq:speed-of-sound-explicit}
\end{align}
and depends on the parameters of the theory as well as the value of the scalar field at the horizon, $\sigma_h$. If the horizon is deep in the IR, the field $\sigma$ grows to be large there and $\sigma_h/\sigma_c \gg 1$. In this limit, eq.~\eqref{eq:speed-of-sound-explicit} simplifies to
\begin{align}
    c_s^2 \:\:&\underset{\sigma_h \gg \sigma_c}{=} \:\: \frac13 - \frac{50}{9 \sigma_c^2}\:.
\end{align}
We indeed see that for large $\sigma_c$, the speed of sound has the expected value $1/\sqrt{3}$ from scaling,\footnote{With no other relevant scales, $s\sim T^3$ in 4 dimensions, so that $c_s^2 = \dd \log T /\dd \log s = 1/3$.} but is smaller for smaller $\sigma_c$, and for some critical value, can become zero. Working with the full $V(\sigma)$, we can plot $c_s^2$ as a function of $\sigma_h$, where $\sigma_h$, the value of the field at the horizon, can be obtained numerically.

\begin{figure}[h!]
    \centering
    \includegraphics[width=0.99\textwidth]{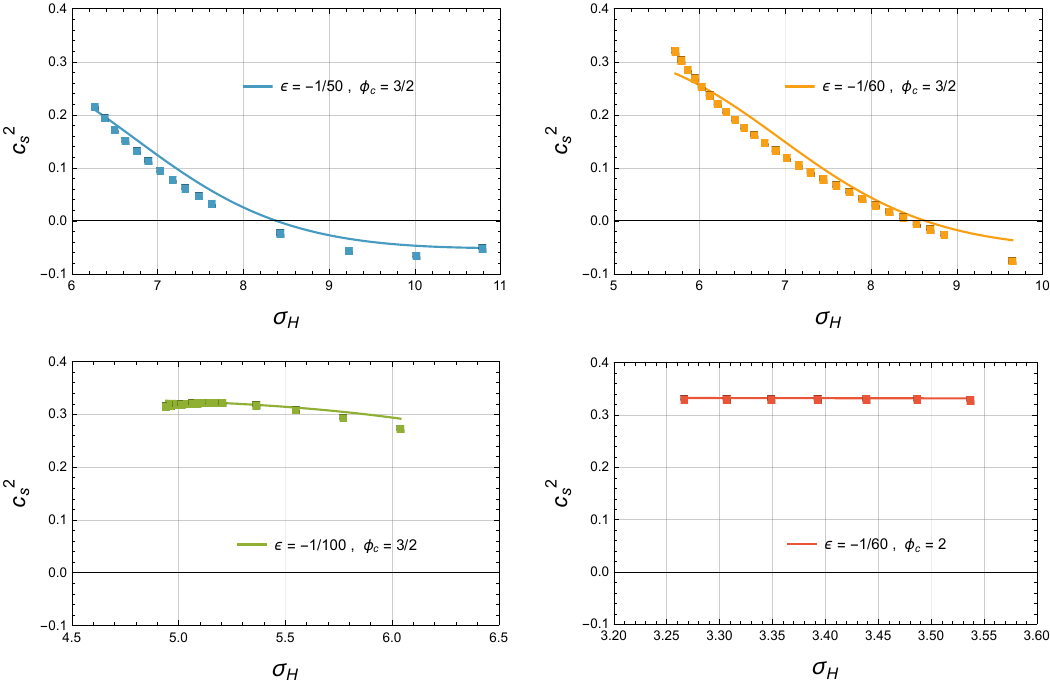}
    \caption{\small{Speed of sound squared for the parameters in eq.~\eqref{eq:params} (param \textbf{A}: top left, param \textbf{B}: top right, param \textbf{C}: bottom left, param \textbf{D}: bottom right). The solid line is obtained using eq.~\eqref{eq:speed-of-sound-explicit}, where $\sigma_h$ is obtained numerically. The dotted lines are from fitting a function to the numerical $(\log T, \log s)$ data and using the exact definition in eq.~\eqref{eq:speed-of-sound-exact-def}.}}
    \label{fig:speed-of-sound}
\end{figure}

Figure~\ref{fig:speed-of-sound} shows the speed of sound squared for the four parameter choices. We calculate it in two different ways---once using the fitted function in eq.~\eqref{eq:FunctionToFit-T-entropy} and using the definition, $c_s^2 \equiv \dd \log T/\dd \log s$, and once using eq.~\eqref{eq:speed-of-sound-explicit} (where $\sigma_h$ is obtained from the numerical solutions).
There are a couple of things to notice in fig.~\ref{fig:speed-of-sound}:
\begin{itemize}
    \item While there is good agreement between the solid lines and the dots, the agreement is only partial for the blue and orange curves. These are the cases with the most back-reaction. This is because the formula for the speed of sound in terms of the potential is only approximate.
    \item For both blue and orange curves, the speed of sound squared crosses zero at some point, and the two methods agree. This happens at larger values of $\sigma_h$, which translates to $\phi_h$ being closer to $\phi_c$. This point is close to where the temperature has a minimum.
    \item The green and red curves are mostly flat, and close to the conformal constant value of $1/3$. Since there is much less back-reaction in these cases, this is in line with the expectation, with the red case being closest to the conformal value. 
\end{itemize}

A match between $c_s^2$ calculated in two different ways gives confidence in the numerical results, and also correlates the minimum temperature with the sound speed approaching zero.

\subsection{$T_\text{min}$ vs $T_c$ and the Absence of Supercooling}
In this section we argue based on our analytical approach that we do not expect a large amount of supercooling in the presence of large back-reaction. Using the analytical expressions derived in this section, and working in $\sigma$ coordinates, we can calculate the value $\sigma_\text{min}$ where the minimum temperature occurs. The existence of a minimum of course requires the scalar contribution to the potential to be important. We can estimate the value $\sigma_\ir$ where the back-reaction from the scalar should be important, which we expect corresponds to the location of the IR brane in the confined geometry, in the absence of any large or small parameters. In this case, the critical temperature $T_c$ is of the same order as the confinement scale which should be approximately $\sigma_\ir$. 

In the limit when the back-reaction is important, all the terms in the expansion in eq.~\eqref{eq:potential_sigma_small} contribute. The potential can be approximated to be
\begin{align}
    V(\sigma) = 2\Lambda + 2\,\tilde{\epsilon}\,\sigma_c^2\,e^{\frac{10}{3}\frac{\sigma}{\sigma_c}}\:,
    \label{eq:Potential-large-sigma-with-const-term}
\end{align}
where $\tilde{\epsilon} = \epsilon\,(\phi_c/\sigma_c)^2$ and $\sigma_c, \phi_c$ are related by eq.~\eqref{eq:phi-vs-sigma}. At $\sigma = \sigma_\ir$, for $\mathcal{O}(1)$ back-reaction, the exponential term is expected to be of the same order as the constant term, giving
\begin{align}
    \sigma_\ir = \frac{3}{10}\sigma_c\,\log\left(\frac{\Lambda}{\tilde{\epsilon}\,\sigma_c^2}\right)\:.
\end{align}
Using the condition $V'/V = \sqrt{2/3}$ for the minimum temperature, we can solve for $\sigma_\text{min}$, which is given as
\begin{align}
    \sigma_\text{min} &= \frac{3}{10}\sigma_c \log \left(\frac{\Lambda}{\tilde{\epsilon}\,\sigma_c^2}\frac{1}{\sigma_*-\sigma_c}\right)\:,\:\: \sigma_* = \sqrt{50/3}\:.
\end{align}
Note that for $\sigma_c > \sigma_* = \sqrt{50/3}$, there is no real solution for $\sigma_\text{min}$, which is consistent with the intuition that for a large $\sigma_c$, equivalent to a large $\phi_c$ by eq.~\eqref{eq:phi-vs-sigma}, the potential is not significantly modified and there is no minimum temperature.

To compare $T_c$ and $T_\text{min}$, we can look at the ratio $\sigma_\ir/\sigma_\text{min}$ which simplifies to
\begin{align}
    \frac{\sigma_\ir}{\sigma_\text{min}} &= \frac{\log\left(\Lambda/(\tilde{\epsilon}\sigma_c)\right) + \log(1/\sigma_c)}{\log\left(\Lambda/(\tilde{\epsilon}\sigma_c)\right) + \log(1/(\sigma_* -\sigma_c))}\:,\:\: \sigma_* = \sqrt{50/3}\:. 
\end{align}
Using the definition in eq.~\eqref{eq:phi-vs-sigma}, we have $\sigma_c > \sqrt{32/3}$. Therefore to have a minimum temperature, $\sigma_c$ must lie in the range $(\sqrt{32/3}, \sqrt{50/3})$. For $\tilde{\epsilon} \ll 1$, $\log(\Lambda/(\tilde{\epsilon}\sigma_c))$ is large and the ratio $\sigma_\ir/\sigma_\text{min}$ is close to 1. For $\sigma_c$ slightly smaller than $\sqrt{50/3}$, $\sigma_\ir/\sigma_\text{min}$ is smaller than 1. For $\sigma_\ir/\sigma_\text{min} \geq 1$, we need $\sigma_c \leq 1/2\sqrt{50/3}$, which is not allowed by eq.~\eqref{eq:phi-vs-sigma}.  Therefore, in the allowed range, $\sigma_\ir/\sigma_\text{min}$ is always smaller than 1, which suggests that the minimum temperature is smaller than the critical temperature. The argument however is only approximate since we did not solve for the correction to the confined phase. 
Note that in the explicit computation~\cite{Buchel:2021yay}, done in the KS geometry, $T_\text{min}$ is indeed smaller than $T_c$, without any parametrically large or small numbers.

\section{Discussion}
\label{sec:discussion}
We set out to find a calculable model with unsuppressed back-reaction in the IR and determine the generality of the supercooled first-order phase transition. Motivated by KKLT, we chose to model the high-temperature dynamics of the warped deformed conifold geometry investigated in ref.~\cite{Buchel:2021yay} by a simplified RS-like setup with an appropriately chosen scalar bulk potential.

\begin{figure}[h!]
    \centering
    \includegraphics[width=0.48\textwidth]{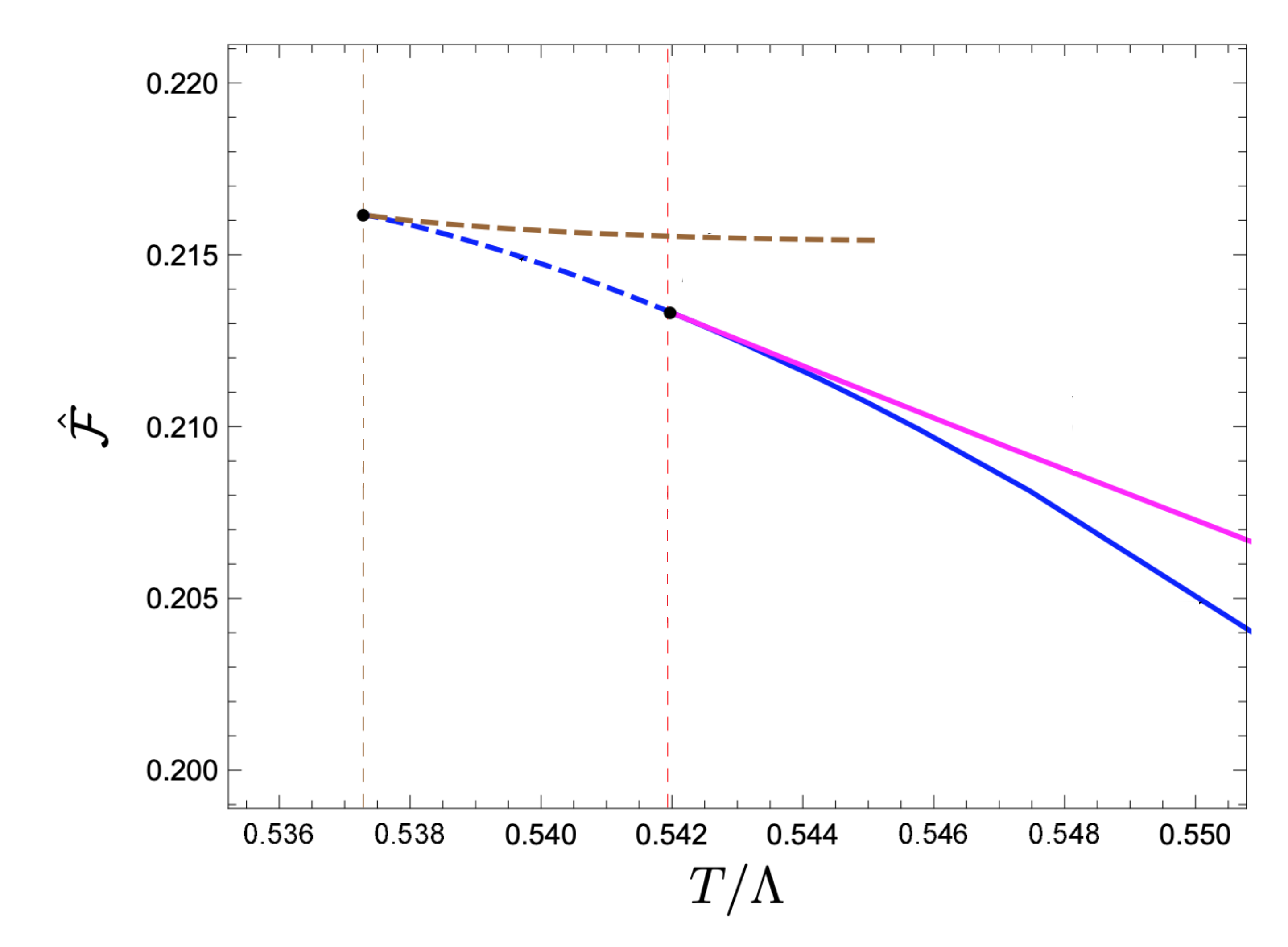}
    \:
    \includegraphics[width=0.48\textwidth]{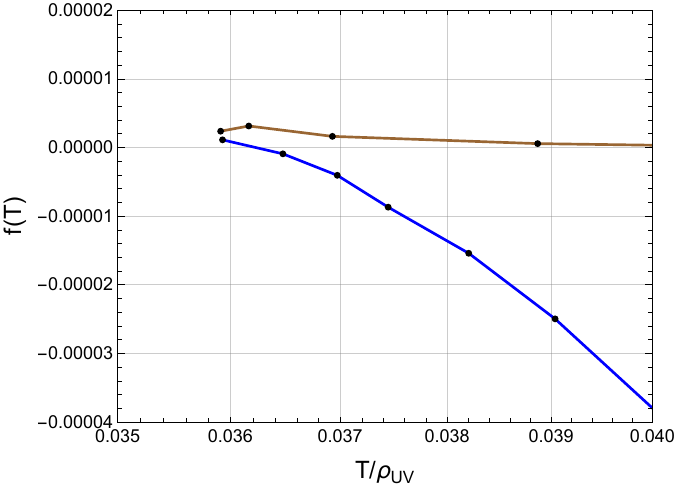}
    \caption{\small{A comparison of the free energy vs temperature curve from ref.~\cite{Buchel:2021yay} (fig. 18 there) and the curve obtained in this work. The magenta curve in the left panel corresponds to deconfined but spontaneously broken chiral symmetric phase, which has no analog in our simplified model. The axes are normalized differently in the two plots.}}
    \label{fig:CompareToBuchel}
\end{figure}

We have found that our model reproduced many of the salient features of the KS deformed conifold model. We found a minimum temperature and hence a free energy that is not single-valued, with the existence of another phase with higher free energy for temperatures higher than the minimum temperature in addition to the usual phase. In fig.~\ref{fig:CompareToBuchel} we show the free energy curve taken from ref.~\cite{Buchel:2021yay} (left panel) alongside the free energy curve we have obtained in this work (right panel) for  parameters corresponding to the parameter set \textbf{A}, also shown in the lower left panel of fig.~\ref{fig:FE-vs-Temperature}. The magenta curve in the left panel of fig.~\ref{fig:CompareToBuchel} corresponds to thermal deconfined states of the cascading gauge theory plasma with spontaneously broken chiral symmetry, and don't have an analog in our simplified setup. However, the existence of minimum temperature and the multi-valued nature of free energy are very clear. Note that the axes are normalized differently in the two plots.
We now briefly consider the generality of our results, the connection to other examples where such features are seen, and the possible dynamics and observational consequences that play out as the temperature changes in the early universe. All of these are targets for future research.

\subsection{Generality of Results}
Even though we focused on a specific potential that was derived to model the reduction of $N$ of the dual theory, our results are expected to hold for general potentials that grow dramatically in the IR. In section~\ref{sec:analyticalEstimates}, we derived analytical formulas for the variation in temperature as a function of horizon location, the speed of sound and the specific heat (eqs.~\eqref{eq:Temperature-vs-horizon-approx}, \eqref{eq:speed-of-sound-approx}, \eqref{eq:specific-heat-approx}). These are all expressed in terms of a general potential $V$ and its derivative $V'$ evaluated at the horizon. In the adiabatic approximation, the minimum temperature, the zero of the speed of sound and the divergence of the specific heat, all occur when $V'/V = \sqrt{2/3}$ (in units of $M_5$). Even for a simple polynomial scalar bulk potential (that grows in magnitude in the IR), this condition will eventually be satisfied (e.g. see~\cite{Bunk:2017fic}). This means that the amount of supercooling is ultimately limited by the dynamics leading to an IR brane.

For a theory that is approximately a CFT in the UV that gets strongly broken in the IR, we expect $V'/V$ (in units of $M_5$) to start with a small value in the UV (where $V$ is approximately constant, dominated by the AdS cosmological constant), and grow to larger values in the IR to eventually be order unity. This should be the case for any theory for which the CFT breaking is sufficiently large in the IR to create a substantial back-reaction that takes the theory out of the perturbative regime (e.g. it is not true for the standard RS stabilized by a GW field with small back-reaction to the geometry). Our results should therefore apply to a wide class of theories in which the IR dynamics is strongly modified. 

In previous works that did not include the back-reaction of the scalar sector on the geometry, there was no way to see a minimum temperature. One needs a feature in the IR in the high temperature phase to provide a scale, which in our case is provided by the scalar potential (there is no IR brane in the deconfined phase). In our work we have included the back-reaction on the geometry from the scalar to incorporate this effect.

\subsection{Connection to Other Examples}
The features seen in our work---the existence of a minimum temperature and of other branches of solutions that are thermodynamically unstable---are seen in other examples. 

The connection to the Klebanov-Strassler (KS) solution~\cite{Klebanov:2000hb} is obvious, and was the motivation for the 5D model we considered. As we have seen, the explicit numerical results in ref.~\cite{Buchel:2021yay} that studied KS black holes demonstrate the existence of a minimum temperature and the multiple-valuedness of entropy and free energy, as we also obtained in our vastly simplified model. Similar features are seen in a 3d analog of KS, as studied in.~\cite{Elander:2020rgv}.

A phase transition in global 5D AdS space in which the field theory lives on $S^3\times R$ also illustrates similar features. This space supports black holes, which dominate the canonical ensemble at high temperatures. There is a transition from thermal AdS (without a black hole) to  AdS space with a black hole (the Schwarzschild AdS solution) at some critical temperature $T_c$: this is the Hawking-Page transition~\cite{Hawking:1982dh}. Crucially, the temperature of the black hole also has a minimum value $T_\text{min}$ which is slightly smaller than $T_c$~\cite{Hawking:1982dh, Yaffe:2017axl}. The entropy and the free energy of the hot phase have a similar shape as we have obtained in this work. The stable branch (with lower free energy) corresponds to a large black hole in AdS (horizon radius is large compared to the AdS radius), while the unstable branch corresponds to a small black hole. The small black hole has lower entropy compared to the large black hole at the same temperature, and has a negative specific heat (like a black hole in flat space). This was also what we found for the unstable branch (see fig.~\ref{fig:entropy-DOF-withBR}). The thermodynamic properties of the black brane on the unstable branch (that appears when the back-reaction is taken into account) are therefore similar to a small black hole in AdS and a flat space black hole. The field theory interpretation and the instabilities present in the small black hole branch is an interesting question and we refer the reader to refs.~\cite{Peet:1998cr, Banks:1998dd, Hubeny:2002xn, Dias:2015pda, Buchel:2015gxa, Dias:2016eto, Yaffe:2017axl} and the references therein. Note that the presence of a minimum temperature requires a non-zero radius $R$ of the $S_3$ on which the field theory lives. The scale $R$ provides an IR scale at which the CFT is explicitly broken.

Another example of a confining gauge theory that has a geometric dual with a shrinking spatial dimension is the Witten model~\cite{Witten:1998zw}. Both the Witten solution~\cite{Witten:1998zw} and the KS solution~\cite{Klebanov:2000hb} are examples where the end of spacetime (dual to confinement of the gauge theory) is modeled by a shrinking spatial circle, $S^2$ for the KS solution and $S^1$ for the Witten solution. In an indirect approach to study confinement, ref.~\cite{Klebanov:2007ws} considered the entanglement entropy of a subregion of the boundary, calculated using the bulk Ryu-Takayanagi (RT) surface in the semi-classical limit. The authors argued that the subregion entropy can be thought of as a free energy, and the length of the subregion can be thought as an inverse temperature. They further showed that there is a maximum length and the subregion entropy as a function of length has the (now) familiar shape of free energy as a function of temperature that we have calculated. 

These discussions indicate that the KS solution, the Witten solution, and  global AdS all show the same general phenomena we have identified in our simple model.

\subsection{Early Universe Dynamics}
\begin{figure}[h!]
    \centering
    \includegraphics[width=0.7\textwidth]{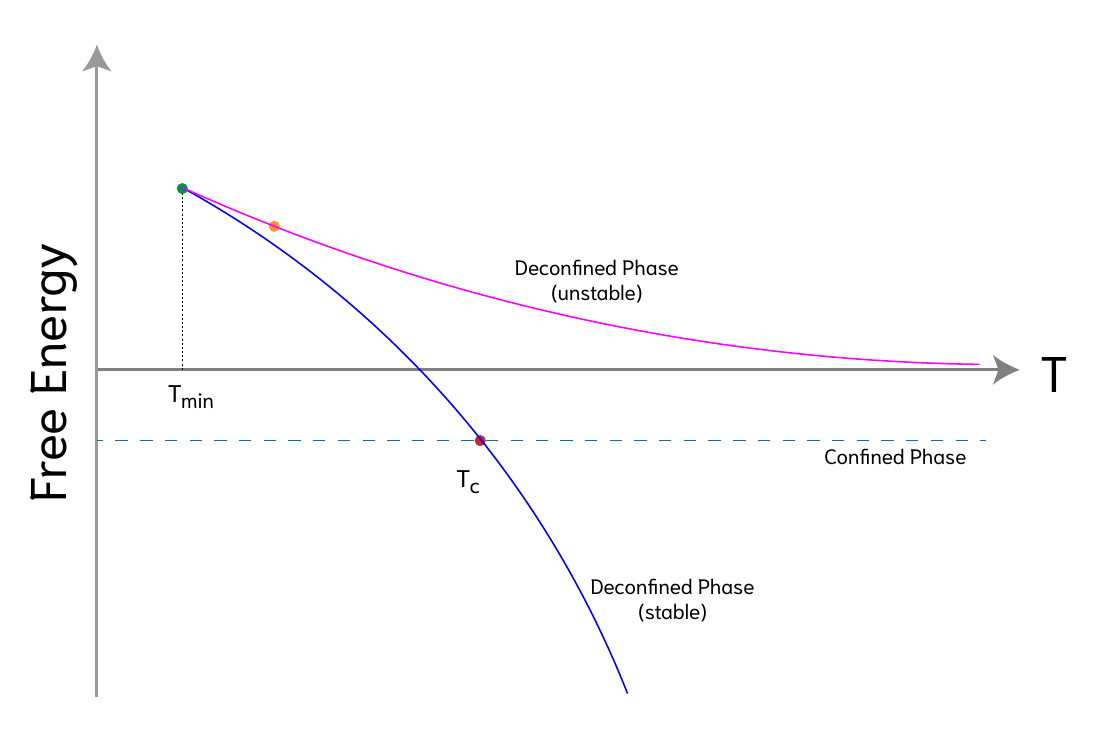}
    \caption{\small{A cartoon of the free energy of the confined phase (dotted line) and deconfined phase (solid line) as a function of temperature. Deconfined phase has a thermodynamically stable branch (in solid blue) and a thermodynamically unstable one (in solid magenta). Red dot shows the critical temperature $T_c$, green dot shows the minimum temperature $T_\text{min}$, and yellow dot indicates possible bifurcation points to other phases.}}
    \label{fig:FE-cartoon}
\end{figure}

Of course we would ultimately like to understand the early universe implications of the results obtained in our work. As we have seen, there are two branches of the black brane phase, and one is always thermodynamically unstable compared to the other stable one. To make the discussions clear, a cartoon of the free energy as a function of temperature, both for the confined phase and for the two branches of the deconfined phase is shown in fig.~\ref{fig:FE-cartoon}. The dashed horizontal line shows the confined phase. The blue and magenta lines show the two branches of the deconfined phase.  

At very high temperatures $T \gg T_c$, the universe is in the thermodynamically preferred deconfined phase (blue line in fig.~\ref{fig:FE-cartoon}). As the temperature redshifts, and becomes smaller than $T_c$, the confined phase is the thermodynamically preferred phase. The vacuum is however stuck in the meta-stable deconfined phase, due to a small rate of transition. As the temperature approaches $T_\text{min}$ from above, the speed of sound squared approaches zero and beyond that point, the black brane phase is not stable against fluctuations.\footnote{A proper analysis would require calculating the quasi-normal modes of the black-brane geometry and checking where they develop an imaginary part (see also ref.~\cite{Buchel:2005nt}).} The magenta curve corresponds to the spinodal region, which is both thermodynamically unstable (since it has a higher free energy) and dynamically unstable (since it has an imaginary speed of sound and negative specific heat). As the temperature approaches the spinodal region (i.e. $T$ approaches $T_\text{min}$ from above), the instabilities  grow and the dynamical evolution becomes complicated. Understanding the time evolution past the minimum temperature requires understanding the dynamics carefully. It is plausible that the upper branch of the black brane phase (magenta curve in fig.~\ref{fig:FE-cartoon}) is never truly populated. Understanding this phase transition better is crucial to understanding any potential gravity wave implications. 

It is plausible that the confined (RS) phase is the only stable phase at low temperatures, and the system will eventually thermalize to that state. Calculating the dynamics of this system is tied to understanding the order of phase transition, and will require a full 5D analysis. 

In the absence of a first-order phase transition, there would be none of the ``conventional'' gravity wave signatures of the RS phase transition. However the spinodal region might itself lead to interesting gravitational wave signals (e.g. see~\cite{Bea:2021zol}). It will be interesting to further explore the early universe signals from the spinodal region.

In our analysis we have assumed a translationally invariant horizon, but it is possible there are more general solutions with different symmetries for the horizon.\footnote{e.g. for $\text{AdS}_5\times S^5$, numerical solutions have been constructed~\cite{Dias:2016eto} where the horizon is localized in the $S^5$ (as opposed to smeared over $S^5$), and corresponds to spontaneous breaking of the dual $SO(6)$ symmetry.} A detailed knowledge of all these phases would be needed to know the precise dynamical evolution of the theory.

\section{Summary and Conclusion}
\label{sec:summary-and-conclusion}
In this work we have investigated the 5D black brane geometry in the presence of a scalar field with a potential motivated by models with a reduction in the local 5D Planck scale seen in higher dimensional supergravity solutions dual to confining gauge theories. In the field theory, this effect models the reduction in the number of colors $N$ as the theory flows from the UV to the IR. Owing to the large potential growth in the Einstein frame there is potentially a large back-reaction on the black brane geometry.

We solved for the fully back-reacted black brane geometry in the presence of a scalar with a potential to model the reduction of $N$, and calculated thermodynamic quantities for the hot phase of the field theory. While many authors have studied the question of phase transition in RS~\cite{Creminelli:2001th, Nardini:2007me, Hassanain:2007js, Konstandin:2010cd, Konstandin:2011dr, Dillon:2017ctw, vonHarling:2017yew,  Bunk:2017fic, Bruggisser:2018mrt, Baratella:2018pxi,  Pomarol:2019aae, Agashe:2019lhy, Fujikura:2019oyi, Baldes:2020kam, Megias:2020vek, Bigazzi:2020phm, Agashe:2020lfz, Agrawal:2021alq, Baldes:2021aph, Csaki:2023pwy, Girmohanta:2023sjv, Eroncel:2023uqf, Mishra:2023kiu, Fichet:2023xbu}, our work is different and focuses on the effect of a strong breaking of the dual CFT in the IR on the deconfined phase (see also ref.~\cite{Megias:2018sxv} for a discussion of reduction of degrees of freedom from back-reaction). In the context of modeling properties of QCD in a holographic setup, similar results have appeared, e.g. see ref.~\cite{Gursoy:2008za} (see also ref.~\cite{Morgante:2022zvc} in the context of GW signal estimation).

In this work we have considered the scenario that the CFT breaking in the IR is large (i.e. there is a large back-reaction in the IR, in the dual 5D picture). In the absence of additional assumptions about the specific CFT dynamics, a large breaking in the IR is the natural expectation (see ref.~\cite{Chacko:2012sy, Bellazzini:2012vz, Coradeschi:2013gda}) for a discussion of this). The interesting thermodynamic features pointed out here emerge for sufficiently large IR breaking, and are also seen in the 10d supergravity examples discussed in sec.~\ref{sec:discussion}, which also suggests that it can be a generic feature in confining theories.

We showed that there are interesting features that emerge in this setup. As the Universe evolves, temperature and entropy can decrease in the black brane phase to a minimum temperature. The effect of a minimum temperature renders the entropy and the free energy of the hot phase significantly different from the no back-reaction limit: the free energy and entropy curves have a turn-around at the minimum temperature and are multiple-valued functions at higher temperatures.\footnote{Our results are consistent with the statements in ref.~\cite{Creminelli:2001th} who argued that the free energy of fields near the black hole horizon redshift to zero and contributes a small amount unless there is a sizeable deformation in the IR.} Beyond the minimum temperature a more detailed study of the dynamics is required but it seems that entropy can decrease (as parameterized by smaller $\alpha$) but this would evolve into the second unstable branch. The dynamics as temperature is lowered might put us on the RS branch but a more detailed study is required.

Our work opens up a number of questions. If the phase structure is as we have argued, what is the dynamics of the phase transition? This question has to be addressed in a fully 5D setup, and an effective treatment in a 4D EFT of the radion, as has been used in literature, will not suffice. This is because we are necessarily in the limit of large back-reaction, and it is not sufficient to consider only the radion at low energy. This question also has to address necessarily the instabilities that the unstable black hole branch (the magenta line in fig.~\ref{fig:FE-cartoon}) is eventually expected to encounter: since the speed of sound crosses zero at the minimum temperature and eventually becomes imaginary, small fluctuations are expected to grow and make the solution unstable. The dynamics question is also tied to understanding the order of the phase transition, which does not proceed by bubble nucleation now presumably. These questions have important observational consequences and we leave them for future work.

\section*{Acknowledgements}
We would like to thank A. Bedroya, I. Bena, A. Buchel, A. Cherman, H. Geng, I. Klebanov, A. Pomarol, M. Reece, M. Strassler and R. Sundrum for useful discussions. We also thank A. Dersy and X. Zhang for useful discussions on the numerical aspects. The work of RKM and LR is supported by the National Science Foundation under Grant Nos. PHY-1620806, PHY-1748958 and PHY-1915071, the Chau Foundation HS Chau postdoc award, the Kavli Foundation grant “Kavli Dream Team”, and the Moore Foundation Award 8342. Part of this work was completed at Aspen Center for Physics, which is supported by National Science Foundation grant PHY-2210452.

\appendix
\section{Potential in the Einstein Frame}
\label{app:pot-in-EF}
One can perform a Weyl rescaling of the metric to transform the action in eq.~\eqref{eq:scalar_action} to a more canonical form with a constant coefficient in front of the Ricci scalar. Consider a rescaling of the metric of the form $g_{MN} \to \Omega(\phi) g_{MN}$. Under this transformation various relevant quantities transform in the following way (specific to 5D) 
\begin{align}
    &g_{\scriptscriptstyle MN} \to \Omega \,g_{\scriptscriptstyle MN}\:,\:\:
    g^{\scriptscriptstyle MN} \to \Omega^{-1} \,g^{\scriptscriptstyle MN}\:,\:\:
    \sqrt{-g} \to \Omega^{5/2}\,\sqrt{-g}\:,\:
    \nonumber \\
    &R \to \frac{1}{\Omega}
    \left(R-4\left(\frac{\Omega''\Omega - \Omega'\,\Omega'}{\Omega^2}(\partial\phi)^2+\frac{\Omega'}{\Omega}D^N\partial_N\phi\right)
    -3\left(\frac{\Omega'}{\Omega}\right)^2(\partial\phi)^2
    \right)\:,
\end{align}
where $\Omega'$ is the derivative w.r.t. its argument $\phi$. Note that in one of the terms, a covariant derivative acts on $\partial\phi$. This is to be kept in mind when integrating this term by parts. It is also useful to keep in mind the identity that the covariant derivative of the metric tensor is identically zero.

Taking $\Omega = (1- \phi/\phi_c )^l$, various terms in the action in eq.~\eqref{eq:scalar_action} transform as 
\begin{align}
    &2M_5^3\int d^5 x\sqrt{-g}\,(1-\phi/\phi_c)^n\,R \nonumber \\
    &\qquad\to 
    2 M_5^3 \int d^5 x\,\sqrt{-g}\,(1-\phi/\phi_c)^{5l/2}\,(1-\phi/\phi_c)^{n}
    \left(1-\phi/\phi_c\right)^{-l}
    \nonumber \\
    &\qquad
    \left(
    R
    +\frac{4l}{\phi_c^2}(1-\phi/\phi_c)^{-2}(\partial\phi)^2
    -\frac{3l^2}{\phi_c^2}(1-\phi/\phi_c)^{-2}(\partial\phi)^2
    +\frac{4l}{\phi_c}(1-\phi/\phi_c)^{-1}(D\partial\phi)
    \right)
    \label{eq:Ricci-after-weyl}
    \\
    &2M_5^3\int d^5 x\sqrt{-g}\,(1-\phi/\phi_c)^m\,(-2\Lambda) \nonumber \\
    &\qquad\to 
    2 M_5^3 \int d^5 x\,\sqrt{-g}\,(1-\phi/\phi_c)^{5l/2}\,(1-\phi/\phi_c)^m\,(-2\Lambda)
    \\
    &2M_5^3\int d^5 x\sqrt{-g}\,\left(-a\,g^{MN}\partial_M\phi\,\partial_N\phi\right) \nonumber \\
    &\qquad\to 
    2 M_5^3 \int d^5 x\,\sqrt{-g}\,(1-\phi/\phi_c)^{5l/2}\,\left(-a(1-\phi/\phi_c)^{-l}\right)(\partial\phi)^2,
    \label{eq:scalar-kinetic-term-after-weyl}
    \\
    &2M_5^3\int d^5 x\sqrt{-g}\,\left(-v(\phi)\right) \nonumber \\
    &\qquad\to 
    2 M_5^3 \int d^5 x\,\sqrt{-g}\,(1-\phi/\phi_c)^{5l/2}\,\left(-v(\phi)\right)\:.
\end{align}
From eq.~\eqref{eq:Ricci-after-weyl} we see that for a $\phi$ independent coefficient of the Ricci scalar $R$ we need
\begin{align}
    3l/2+n=0\:.
\end{align}
After the Weyl rescaling, and after integration by parts,\footnote{The boundary term coming from integration by parts cancels the boundary term with extrinsic curvature. The extrinsic curvature term ensures that variation in fields give the equations of motion without any boundary terms, and a Weyl transformation is a special kind of variation in metric, proportional to metric itself.} the kinetic term for the scalar in eq.~\eqref{eq:scalar-kinetic-term-after-weyl} is given as (using $n=-3l/2$)
\begin{align}
    2M_5^3\int d^5 x\,\sqrt{-g}
    \left(
    -a\left(1-\frac{\phi}{\phi_c}\right)^{-n} 
    -\frac{4n^2}{3\phi_c^2} \left(1-\frac{\phi}{\phi_c}\right)^{-2}
    \right)
    (\partial\phi)^2\:.
\end{align}
from which one can read off $G(\phi)$ appearing in eq.~\eqref{eq:scalar_action_after_weyl_rescaling}:
\begin{align}
    G(\phi) = 2a\left(1-\frac{\phi}{\phi_c}\right)^{-n} + \frac{8n^2}{3\phi_c^2}\left(1-\frac{\phi}{\phi_c}\right)^{-2}\:.
\end{align}
To obtain the potential after the weyl rescaling, we can look at the non-derivative part of the transformed action:
\begin{align}
    2M_5^3\int d^5 x\sqrt{-g}(1-\phi/\phi_c)^{5l/2}(-v(\phi)-2(1-\phi/\phi_c)^m\Lambda)\:,
\end{align}
from which the potential (that includes the cosmological constant) can be read off, and is given as (using $n=-3l/2$)
\begin{align}
    V(\phi) = v(\phi)\left(1-\frac{\phi}{\phi_c}\right)^{-\frac{5n}{3}} 
    + 2\Lambda\left(1-\frac{\phi}{\phi_c}\right)^{-\frac{5n}{3}+m}\:.
\end{align}

\section{IR Expansion and Regularity at the Horizon}
\label{app:IR-derivatives}
Here we give the expression for the functions $a_1$ and $\phi_1$ that enter in the regularity condition in the IR. One can obtain an expression for them without specifying the functions $V(\phi), G(\phi)$. We substitute $\phi(\xi) = \phi_0 + \phi_1\,\xi$, $a(\xi) = a_0 + a_1\,\xi$ in the equations of motion and solve for the coefficients $a_1, \phi_1$, keeping to zeroth order in $\xi$. The resulting expressions are
\begin{align}
    a_1(a_0, \phi_0) &= \frac{a_0}{32\alpha}\left(\frac{3}{G(\phi_0)}\left(\frac{V'(\phi_0)}{V(\phi_0)}\right)^2-8\:\right),
    \nonumber \\
    \phi_1(a_0, \phi_0) &= -\frac{3}{4\alpha}\frac{1}{G(\phi_0)} \frac{V'(\phi_0)}{V(\phi_0)}\:.
\end{align}
Given an explicit form of $V(\phi)$ and $G(\phi)$, these functions can be readily calculated. Given $G(\phi)$ and $V(\phi)$ in eq.~\eqref{eq:G-and-V-specific}, the functions are given as
\begin{align}
    a_1 (a_0, \phi_0) &= -\frac{a_0}{16\alpha}
    \left(
    4 - \frac{\epsilon^2\phi_0^2(3+2\phi_0/\phi_c)^2}{\left(3+16/\phi_c^2\right)\left(\Lambda  \left(1-\phi_0/\phi_c\right)^{10/3}+\epsilon \phi_0^2\right)^2}
    \right)\:,
    \nonumber \\
    \phi_1 (a_0, \phi_0) &= - \frac{3\epsilon}{4\alpha}\frac{\phi_0}{\phi_c}\frac{(1-\phi_0/\phi_c) (3+ 2\phi_0/\phi_c)}{  \left(3+16/\phi_c^2\right) \left(\Lambda  \left(1-\phi_0/\phi_c\right)^{10/3}+\epsilon \phi_0^2\right)}\:.
\end{align}

\section{Analytical Results}
\label{app:AnalyticalResults}
Starting with the metric
\begin{align}
    \dd s^2 &= e^{2 A(\sigma)}\left(-f(\sigma)\,\dd t^2 + \dd x^2\right) + e^{2B(\sigma)}\frac{\dd \sigma^2}{f(\sigma)}\:, \qquad \sigma_\uv \leq \sigma \leq \sigma_h\:,
    \label{eq:metric-sigma-coordinates-app}
\end{align}
the equations of motion are given as ($' \equiv \partial_\sigma$)
\begin{align}
    A'' - A' B' + 1/6 & = 0\:,
    \label{eq:EOM-sigmaCoordinates-1}
    \\
    f'' + (4A'-B')f' &= 0\:,
    \label{eq:EOM-sigmaCoordinates-2}
    \\
    6A'f' + f(24A'^2-1) + 2 e^{2B}V &= 0\:,
    \label{eq:EOM-sigmaCoordinates-3}
    \\
    f(4A' - B') + f' - e^{2B}V' &=0\:.
    \label{eq:EOM-sigmaCoordinates-4}
\end{align}
Eqn.~\eqref{eq:EOM-sigmaCoordinates-4} is redundant, and can be obtained by differentiating eq.~\eqref{eq:EOM-sigmaCoordinates-3} and using eqns.~\eqref{eq:EOM-sigmaCoordinates-1},~\eqref{eq:EOM-sigmaCoordinates-2}. Using $f(\sigma_h) = 0$, eqn.~\eqref{eq:EOM-sigmaCoordinates-3},~\eqref{eq:EOM-sigmaCoordinates-4} evaluated at $\sigma = \sigma_h$ give
\begin{align}
    f'(\sigma_h) &= -\frac{e^{2B(\sigma_h)}}{3A'(\sigma_h)} V(\sigma_h)\:,
    \label{eq:f-prime-1}
    \\
    f'(\sigma_h) & = e^{2B(\sigma_h)} V'(\sigma_h)\:,
    \label{eq:f-prime-2}
\end{align}
equating eqs.~\eqref{eq:f-prime-1} and~\eqref{eq:f-prime-2}, one can obtain the relation
\begin{align}
    A'(\sigma_h) &=-\frac13\frac{V(\sigma_h)}{V'(\sigma_h)}\:.
    \label{eq:A-prime-in-terms-of-Potential}
\end{align}
Using eq.~\eqref{eq:f-prime-1} and the definition of temperature in terms of $f'$ from eq.~\eqref{eq:Temperature-sigma-coordinates}, we have
\begin{align}
    T &= \frac{1}{12\pi}e^{A(\sigma_h)} e^{B(\sigma_h)}\frac{V(\sigma_h)}{A'(\sigma_h)}\:.
    \label{eq:T-FullEqn-app}
\end{align}
Taking log and differentiating eq.~\eqref{eq:T-FullEqn-app} w.r.t $\sigma_h$ we obtain
\begin{align}
    \frac{\dd \log T}{\dd \sigma_h}
    &= 
    A'(\sigma_h) + B'(\sigma_h) + \frac{V'(\sigma_h)}{V(\sigma_h)} - \frac{A''(\sigma_h)}{A'(\sigma_h)}
    \nonumber \\
    &\qquad \frac{\partial}{\partial \sigma_h}\left(A + B + \log V - \log A'(\sigma_h)\right)\:.
    \label{eq:T-differentiate-full}
\end{align}
Note that we have been careful about $\dd/\dd \sigma_h$. The temperature $T$ depends on $\sigma_h$ in two ways: through the functions $A, B, V$ evaluated at $\sigma = \sigma_h$ and through explicit dependence on $\sigma_h$. The former gives the first line in eq.~\eqref{eq:T-differentiate-full} and the latter gives the second line in eq.~\eqref{eq:T-differentiate-full}.

The $\partial/\partial \sigma_h$ terms are dropped in the so-called ``adiabatic approximation''~\cite{Gubser:2008ny}. Using the equations of motion~\eqref{eq:EOM-sigmaCoordinates-1}, we can eliminate $A''$ from eq.~\eqref{eq:T-differentiate-full} to obtain
\begin{align}
    \frac{\dd \log T}{\dd \sigma_h} 
    & = 
    A'(\sigma_h) + \frac{V'(\sigma_h)}{V(\sigma_h)} + \frac{1}{6A'(\sigma_h)}\:\:+ \:\:\partial_{\sigma_h}\text{ terms}\:.
    \label{eq:T-differentiate-adiabatic-intermediate}
\end{align}
Now using the earlier derived relation between $A'$ and the potential $V$ and $V'$ in eq.~\eqref{eq:A-prime-in-terms-of-Potential} we get
\begin{align}
    \frac{\dd \log T}{\dd \sigma_h} 
    &= -\frac{V(\sigma_h)}{V'(\sigma_h)}\left(\frac13 - \frac12 \left(\frac{V'(\sigma_h)}{V(\sigma_h)}\right)^2\right) \:\:+ \:\:\partial_{\sigma_h}\text{ terms}\:. 
    \label{eq:T-differentiate-adiabatic-final}
\end{align}

Next, using the expression for entropy density $s$, and using eq.~\eqref{eq:A-prime-in-terms-of-Potential} again, we obtain
\begin{align}
    \frac{\dd \log s}{\dd \sigma_h} &= -\frac{V'(\sigma_h)}{V(\sigma_h}  \:\:+ \:\:\partial_{\sigma_h}\text{ terms}\:.
    \label{eq:Entropy-differentiate-adiabatic-final}
\end{align}

Using eqs.~\eqref{eq:T-differentiate-adiabatic-final},~\eqref{eq:Entropy-differentiate-adiabatic-final}, we finally obtain
\begin{align}
    c_s^2 
    &= \frac13 - \frac12\left(\frac{V'(\sigma_h)}{V(\sigma_h)}\right)^2  \:\:+ \:\:\partial_{\sigma_h}\text{ terms}\:.
    \label{eq:SpeedOfSound-adiabatic}
    \\
    C_V &= 8\pi M_5^3 e^{3A(\sigma_h)}\left( \frac13 - \frac12\left(\frac{V'(\sigma_h)}{V(\sigma_h)}\right)^2 \right)^{-1}
    \:\:+ \:\:\partial_{\sigma_h}\text{ terms}\:.
    \label{eq:SpecificHeat-Adiabatic}
\end{align}

Finally, we obtain
\begin{align}
    \frac{\dd \log (s/T^3)}{\dd \sigma_h} &= -\frac32\frac{V'(\sigma_h)}{V(\sigma_h)}\:\:+ \:\:\partial_{\sigma_h}\text{ terms}\:,
\end{align}
integrating which, we get $s/T^3 \propto \left|V(\sigma_h)\right|^{-3/2}$, in the adiabatic approximation. Since $s/T^3$ is a measure of degrees of freedom, this shows that as the potential gets large in the IR, the degrees of freedom reduce. This is very much the intuition from a dual viewpoint. 
\section{Alternate Numerical Method}
\label{app:numericalMethod-Alternate}

In this section we present an alternative numerical method to obtain the metric and the scalar profile, working in a gauge where the radial direction is chosen to be the scalar $\sigma$ itself (henceforth called the scalar gauge). Our steps closely follow those in ref.~\cite{Gubser:2008ny}. 

In the scalar gauge, the metric is written in eq.~\eqref{eq:metric-sigma-coordinates-app} and the equations of motion are written in eqs.~\eqref{eq:EOM-sigmaCoordinates-1},\eqref{eq:EOM-sigmaCoordinates-2},\eqref{eq:EOM-sigmaCoordinates-3},\eqref{eq:EOM-sigmaCoordinates-4}. These equations of motion enjoy a weak form of integrability in the sense that if a smooth function $G(\sigma) = A'(\sigma)$ is specified, the other unknown functions can be solved in terms of $G, G'$ and the potential is also fixed in terms of $G, G'$. 

To be more concrete, taking $A'(\sigma) = G(\sigma)$, the solution for $A$ is trivial:
\begin{align}
    A(\sigma) &= A_0 + \int_{\sigma_0}^\sigma \dd x\, G(x)\:,
    \label{eq:sol-A}
\end{align} 
where $A_0 = A(\sigma_0)$ and $\sigma_0$ is arbitrary at this point. Using eq.~\eqref{eq:EOM-sigmaCoordinates-1}, one can solve for $B$ next:
\begin{align}
    B(\sigma) &= B_0 + \int_{\sigma_0}^\sigma \dd x\, \frac{G'(x)+1/6}{G(x)}\:,
    \label{eq:sol-B}
\end{align}
where $B_0 = B(\sigma_0)$. Next, using the obtained solution for $A(\sigma), B(\sigma)$ one can use eq.~\eqref{eq:EOM-sigmaCoordinates-2} to solve for $f(\sigma)$:
\begin{align}
    f(\sigma) &= f_0 + f_1 \int_{\sigma_0}^\sigma \dd x\, e^{-4A(\sigma)+B(\sigma)}\:,
    \label{eq:sol-f}
\end{align}
where $f_0, f_1$ are integration constants. Finally, using eq.~\eqref{eq:EOM-sigmaCoordinates-3}, the potential $V(\sigma)$ is related to the solved quantities:
\begin{align}
    V(\sigma) &= \frac{1}{2}e^{-2B(\sigma)}f(\sigma)\left(1-24 G(\sigma)^2 - 6 G(\sigma) \frac{f'(\sigma)}{f(\sigma)}\right)\:.
    \label{eq:sol-V}
\end{align}

Combining eqs.~\eqref{eq:sol-A},~\eqref{eq:sol-B},~\eqref{eq:sol-f} and~\eqref{eq:sol-V} one can obtain a single non-linear second order differential equation for $G(\sigma)$:
\begin{align}
    \frac{G'}{G+V/3V'} &= \frac{\dd}{\dd \sigma} \log \left(\frac{G'}{G} + \frac{1}{6G}-4G-\frac{G'}{G+V/3V'}\right)\:.
    \label{eq:non-linear-eqn-G}
\end{align}
One special feature of the equation is that starting with the value of the scalar at the horizon, $\sigma = \sigma_h$, one can find a series solution to the equation around $\sigma = \sigma_h$, and all the coefficients in this series solution are fully fixed in terms of the potential $V(\sigma)$ and its derivatives. The solution up to a few terms looks like
\begin{align}
    G(\sigma) &= G_0 + G_1\left(\sigma-\sigma_h\right) + G_2\left(\sigma-\sigma_h\right)^2 + \cdots\: ,
    \\
    G_0 &= -\frac{1}{3}\frac{V(\sigma_h)}{V'(\sigma_h)} \\
    G_1 &= \frac{1}{6}\left(\frac{V(\sigma_h)V''(\sigma_h)}{V'(\sigma_h)^2}-1\right) \\
    G_2 &= \frac{V(\sigma_h)}{27 V'(\sigma_h)}+\frac{V'''(\sigma_h) V(\sigma_h)}{18 V'(\sigma_h)^2}-\frac{V(\sigma_h)^2 V''(\sigma_h)}{27 V'(\sigma_h)^3}-\frac{7 V(\sigma_h) V''(\sigma_h)^2}{72 V'(\sigma_h)^3}+\frac{V''(\sigma_h)}{24 V'(\sigma_h)} \:.
\end{align}
Note that at $\sigma = \sigma_h$, $G+V/3V' = 0$ and the LHS of eq.~\eqref{eq:non-linear-eqn-G} is infinite, so that a direct numerical solution to eq.~\eqref{eq:non-linear-eqn-G} is not possible. The way out is to first obtain the series solution to a high order, use that to obtain $G(\sigma = \sigma_h - \delta), G'(\sigma = \sigma_h - \delta)$ for very small positive $\delta$, and then solve from $\sigma = \sigma_h - \delta$ all the way to $\sigma = \sigma_\uv$ using $G(\sigma = \sigma_h - \delta), G'(\sigma = \sigma_h - \delta)$ as the initial conditions.  Interestingly, the need to regulate the location of the horizon for a numerical solution was also seen in the $\xi$ coordinates used in the main text.

After obtaining $G(\sigma)$, one still needs to obtain the integration constants to fix the functions $A(\sigma), B(\sigma)$ uniquely. These integration constants are obtained by requiring that for very small values of $\sigma$ (i.e. close to UV), the potential $V(\sigma)$ can be approximated by $2\epsilon \sigma^2$, for which the scalar solution is known analytically, and $A(\sigma)$ is approximately the same as no back-reaction case. After these steps, the entropy and the temperature are uniquely fixed and are given as (further specific details can be found in ref.~\cite{Gubser:2008ny}):
\begin{align}
    s &= 8 \pi M_5^3\,\sigma_h^{3/\epsilon}\,\exp \left(3\int_{\sigma_\uv}^{\sigma_h}\dd \sigma \left(G(\sigma) - \frac{1}{\epsilon\,\sigma}\right)\right)\:,
    \\
    T &= \frac{\sigma_h^{1/\epsilon}}{\pi}\,\frac{V(\sigma_h)}{V(\sigma_\uv)}\,\exp\left(\int_{\sigma_\uv}^{\sigma_h} \dd \sigma\left(G(\sigma) - \frac{1}{\epsilon\,\sigma} + \frac{1}{6\,G(\sigma)}\right)\right)\:.
\end{align}
These expressions are derived in the limit of $|\epsilon|\ll 1$. We choose $\phi_\uv = 10^{-2}$, which fixes $\sigma_\uv$, given $\phi_c$, using eq.~\eqref{eq:phi-vs-sigma}. Using these expressions one can obtain $s(\sigma_h), T(\sigma_h)$ and therefore $s(T)$ parametrically. Finally, using eq.~\eqref{eq:FE-from-Entropy} and the procedure discussed in sec.~\ref{sec:NumericalResults-FE}, the free energy as a function of temperature can be obtained.

\begin{figure}[h!]
    \centering
    \includegraphics[width=0.9\textwidth]{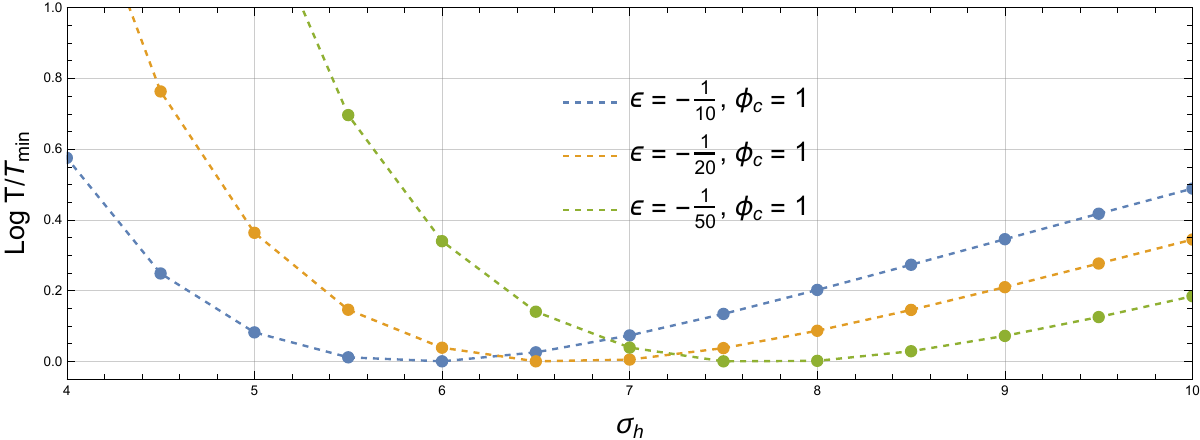}
    \caption{\small{Temperature (normalized by minimum Temperature) as a function of $\sigma_h$ for three parameter choices: $\epsilon = -1/10$ (blue), $\epsilon = -1/20$ (orange) and $\epsilon = -1/50$ (green). For all these cases, $\phi_c = 1$, and $\phi_\uv \approx \sigma_\uv \approx 10^{-2}$. Dots are numerically obtained points, joined by a smooth dashed line.}}
    \label{fig:T-min-vs-eps}
\end{figure}

\begin{figure}[h!]
    \centering
    \includegraphics[width=0.47\textwidth]{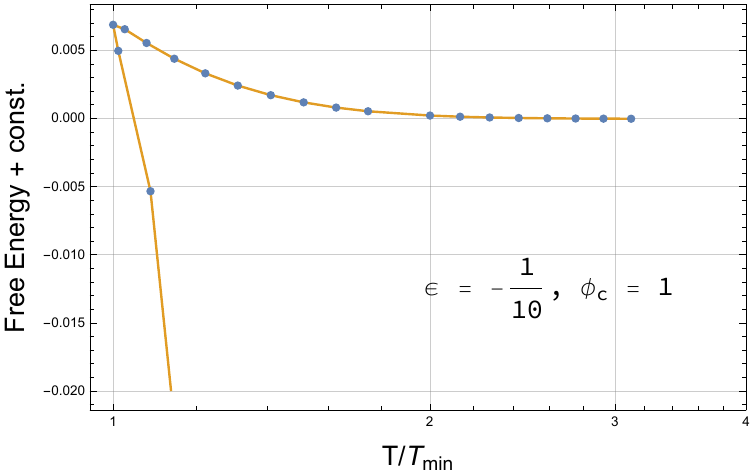}
    \:\:\:
    \includegraphics[width=0.47\textwidth]{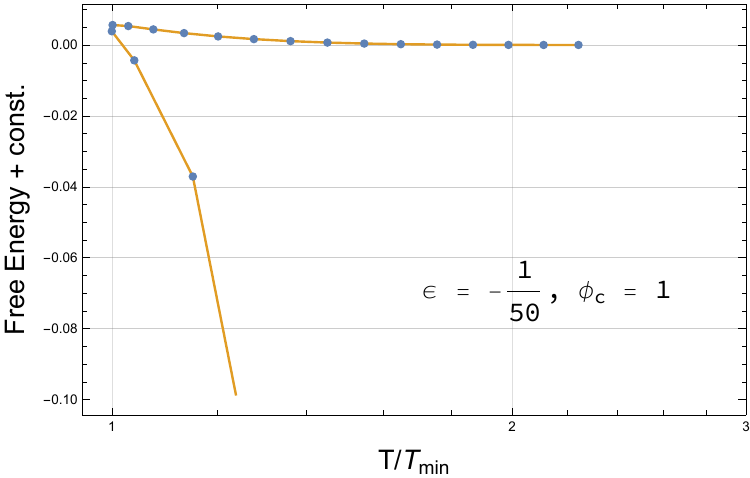}
    \caption{\small{Free energy as a function of Temperature, for two choice of parameters: $\epsilon = -1/10, \phi_c = 1$ (left) and $\epsilon = -1/50, \phi_c = 1$ (right). For both cases, $\phi_\uv \approx \sigma_\uv \approx 10^{-2}$. Dots are numerically obtained points, joined by a smooth line. $M_5$ has been set to 1 in all these plots.}}
    \label{fig:FE-vs-T}
\end{figure}

Fig.~\ref{fig:T-min-vs-eps} shows Temperature (normalized by the corresponding minimum temperature) as a function of $\sigma_h$, for fixed $\phi_c = 1$, as $\epsilon$ is varied. As the magnitude of $\epsilon$ reduces, the value of $\sigma_h$ where the minimum temperature occurs increases. Fig.~\ref{fig:FE-vs-T} shows free energy as a function of temperature, for two choices of parameters. The fin-like feature is seen clearly for both the parameter choices.  

\newpage
\bibliographystyle{utphys}
\bibliography{references}
\end{document}